\documentclass[a4paper,fleqn,usenatbib]{mnras}

\usepackage{newtxtext,newtxmath}

\usepackage[T1]{fontenc}
\usepackage{ae,aecompl}

\usepackage{graphicx}   
\usepackage{amsmath}    
\usepackage{amssymb}    
\usepackage{multirow}
\usepackage[bigfiles]{media9}
\usepackage{pdflscape}

\newcommand{\hi}{H\,{\sc i}}
\newcommand{\HI}{H\,{\sc i}}

\newcommand{\kms}{km\,s$^{-1}$}
\newcommand{\msun}{M$_{\odot}$}

\title[CHILES: \HI\ at $z=0.12$ and $z=0.17$]{CHILES: \HI\ morphology and galaxy environment at z=0.12 and z=0.17}

\author[K. M. Hess et al.]{Kelley M.~Hess$^{1,2,3}$\thanks{E-mail: hess@astro.rug.nl}, 
Nicholas M.~Luber$^{4}$, Ximena Fern{\'a}ndez$^{5}$, Hansung B.~Gim$^{6,7}$, \newauthor
J.~H.~van Gorkom$^{4}$, Emmanuel Momjian$^{8}$, Julia Gross$^{4}$, Martin Meyer$^{3,9}$, \newauthor 
Attila Popping$^{3,9}$, Luke J.~M.~Davies$^{9}$, Lucas Hunt$^{10,11,12}$, Kathryn Kreckel$^{13}$, \newauthor 
Danielle Lucero$^{14,1}$, D.~J.~Pisano$^{10,15,16}$, Monica Sanchez-Barrantes$^{8,17}$, Min S.~Yun$^{6}$, \newauthor 
Richard Dodson$^{9}$, Kevin Vinsen$^{9}$, Andreas Wicenec$^{3,9}$, Chen Wu$^{9}$, \newauthor 
Matthew A.~Bershady$^{18}$, Aeree Chung$^{19}$, Julie D.~Davis$^{18}$, Jennifer Donovan Meyer$^{20}$, \newauthor 
Patricia Henning$^{17}$, Natasha Maddox$^{2}$, Evan T.~Smith$^{10,15}$, J.~M.~van der Hulst$^{1}$, \newauthor
Marc A.~W.~Verheijen$^{1}$, Eric M.~Wilcots$^{18}$
\\
$^{1}$Kapteyn Astronomical Institute, University of Groningen, Landleven 12, 9747 AD, Groningen, The Netherlands\\
$^{2}$ASTRON, the Netherlands Institute for Radio Astronomy, Postbus 2, 7990 AA, Dwingeloo, The Netherlands\\
$^{3}$Australian Research Council, Centre of Excellence for All-sky Astrophysics (CAASTRO), Australia\\
$^{4}$Department of Astronomy, Columbia University, New York, NY 10027, USA\\
$^{5}$Department of Physics and Astronomy, Rutgers, The State University of New Jersey, Piscataway, NJ 08854-8019, USA\\
$^{6}$Department of Astronomy, University of Massachusetts, Amherst, MA 01003, USA\\
$^{7}$School of Earth and Space Exploration, Arizona State University, Tempe, AZ 85281, USA\\
$^{8}$National Radio Astronomy Observatory, P.O. Box 0, Socorro, NM 87801, USA\\
$^{9}$International Centre for Radio Astronomy Research, The University of Western Australia, Crawley, WA 6009, Australia\\
$^{10}$Department of Physics and Astronomy, West Virginia University, P.O. Box 6315, Morgantown, WV 26506, USA\\
$^{11}$U.S.~Naval Observatory, 3450 Massachusetts Avenue NW, Washington, DC 20392, USA\\
$^{12}$George Mason University, Department of Physics \& Astronomy, 4400 University Drive, Fairfax, VA 22030, USA\\
$^{13}$Max Planck Institute for Astronomy, Knigstuhl 17, D-69117 Heidelberg, Germany\\
$^{14}$Department of Physics, Virginia Tech, 850 West Campus Drive, Blacksburg, VA 24061\\
$^{15}$Center for Gravitational Waves and Cosmology, West Virginia University, Chestnut Ridge Research Building, Morgantown, WV 26505\\
$^{16}$Adjunct Astronomer at Green Bank Observatory, Green Bank, WV, USA\\
$^{17}$Department of Physics and Astronomy, University of New Mexico, Albuquerque, NM 87131, USA\\
$^{18}$Department of Astronomy, University of Wisconsin-Madison, Madison, WI 53706, USA \\
$^{19}$Department of Astronomy, Yonsei University, 50 Yonsei-ro, Seodaemun-gu, Seoul 03722, Korea\\
$^{20}$National Radio Astronomy Observatory, Charlottesville, VA 22901, USA
}

\date{Accepted XXX. Received YYY; in original form ZZZ}

\pubyear{2017}

\begin{document}
\label{firstpage}
\pagerange{\pageref{firstpage}--\pageref{lastpage}}
\maketitle

\begin{abstract}

We present a study of 16 \hi-detected galaxies found in 178 hours of observations from Epoch 1 of the COSMOS \HI\ Large Extragalactic Survey (CHILES).  We focus on two redshift ranges between $0.108\le z\le0.127$ and $0.162\le z\le0.183$ which are among the worst affected by radio frequency interference (RFI).  While this represents only 10\% of the total frequency coverage and 18\% of the total expected time on source compared to what will be the full CHILES survey, we demonstrate that our data reduction pipeline recovers high quality data even in regions severely impacted by RFI.  We report on our in-depth testing of an automated spectral line source finder to produce \hi\ total intensity maps which we present side-by-side with significance maps to evaluate the reliability of the morphology recovered by the source finder.  We recommend that this become a common place manner of presenting data from upcoming \hi\ surveys of resolved objects.
We use the COSMOS 20k group catalogue, and we extract filamentary structure using the topological DisPerSE algorithm to evaluate the \hi\ morphology in the context of both local and large-scale environments and we discuss the shortcomings of both methods.  Many of the detections show disturbed \HI\ morphologies suggesting they have undergone a recent interaction which is not evident from deep optical imaging alone.  Overall, the sample showcases the broad range of ways in which galaxies interact with their environment.  This is a first look at the population of galaxies and their local and large-scale environments observed in \HI\ by CHILES at redshifts beyond the $z=0.1$ Universe.

\end{abstract}


%
%
%

\section{Introduction}

Neutral atomic hydrogen (\HI) is the state through which all baryonic matter must pass on its way from the filaments of large scale structure and galaxy halos to galaxy disks, and via molecular clouds to stars.  How galaxies acquire their gas from the environment, and the degree to which the environment triggers gas loss through gravitational or hydrodynamic interactions is an open question.  Observations show that \HI\ is removed from galaxies through various processes across a range of environments from galaxy clusters to groups (e.g.~\citealt{BravoAlfaro00,VerdesMontenegro01,Chung09,Hess13,Jaffe15,Jaffe16,Denes16}), while simulations suggest that cold gas flowing through filaments is important for refueling galaxies at all times over the history of the Universe (\citealt{Sancisi08,Keres09,Putman12,Spring17}).  Ultimately, the total gas content of a galaxy is the sum of these competing activities, and understanding the balance of these processes is important for interpreting the nature of galaxies.  The availability or lack of a cold gas reservoir impacts star formation and the subsequent evolution of stellar populations in galaxies (e.g.~\citealt{Schawinski14}).  

The ability to study the impact of environment on the cold gas content of galaxies has grown substantially with the relatively recent availability of wide-area \HI\ 21 cm surveys such as the \HI\ Parkes All Sky Survey, (HIPASS; \citealt{Meyer04}) and Arecibo Legacy Fast ALFA (ALFALFA) survey (\citealt{Giovanelli05,Haynes18}).
However, as described below, the variation of \HI\ gas content observed across different environments presents a complicated picture, which in part is driven by different definitions of environment.

The GALEX Arecibo SDSS Survey (GASS) found that high stellar mass galaxies ($M_*>10^{10}$~\msun) in groups and clusters have less cool gas than those in low density environments \citep{Catinella13}, while HIPASS found that high stellar mass galaxies are gas rich when they live in filaments as compared to voids \citep{Kleiner17}.  The interpretation may be that filaments provide a large reservoir from which gas cools and accretes onto galaxies as compared to voids, but groups and clusters represent higher density knots within the filaments where gas is removed or processed through interactions or star formation. 

On the low stellar mass end, the Extended GASS survey (xGASS) found that galaxies in very small groups ($N=2$ -- 3) have higher gas fractions than those in isolation \citep{Janowiecki17}, while HIPASS found no difference in environmental dependence between filaments and voids \citep{Kleiner17}.  In this case, small groups may be surrounded by a relatively untapped gas reservoir or be slow to use their gas, while filaments encompass such a range of variation within them (both small and large groups, and non-group galaxies) that such a definition of environment smoothes over the details of what is going on within it.  

In more massive groups and clusters low \hi\ mass galaxies are lacking relative to the field as evidenced by a shallower low mass slope in the group/cluster \hi\ mass function (HIMF; \citealt{Verheijen00,Rosenberg02,deBlok02,Freeland09,Pisano11}), and there is evidence that low mass galaxies preferentially lose their gas before more massive galaxies do in these environments (e.g.~\citealt{Hess13,Odekon16}). Low \hi\ mass galaxies tend to also have low stellar mass, so it may be that their gas is less tightly bound, and more readily lost in the group and cluster environment.

Clearly, careful consideration of the definition of environment in these studies is critical to understanding the compiled results. The results demonstrate that using multiple methods to define environment is valuable for disentangling the whole picture of galaxy evolution (e.g.~\citealt{CroneOdekon18}), and a single environmental definition may only be effective at picking out a limited range of structures.  

Multiple methods exist for defining galaxy environment: counting galaxies within constant apertures and density estimates based on nearest neighbors \citep{Jones16}; friends-of-friends algorithms \citep{Berlind06,Tempel12} and iterative halo finders \citep{Yang07,Lim17} to identify common parent dark matter halos; and tessellation algorithms to identify connected networks of galaxies (e.g.~\citealt{Springob07}), to name a few.  
The strength with which correlations are found may depend on whether groups are optically selected \citep{Catinella13,Hess13}, or X-ray selected \citep{Odekon16}.  

Stellar mass is the most reliable tracer of dark matter halo mass (e.g.~\citealt{2017arXiv170703406P}), which is arguably the most fundamental parameter by which to define environment.  Thus, in order to accurately measure environment, \HI\ observations rely on complementary optical imaging and spectroscopic surveys.  By comparison, \HI-detected objects, as a population, are one of the least well clustered subsets of galaxies (e.g.~\citealt{Meyer07,Papastergis13}).
For this study we use the existing friends-of-friends group finder that was run on the zCOSMOS 20k data \citep{Knobel12}, and run a discrete persistent source extractor to find filaments using optical spectroscopic redshifts (N.~Luber et al., in preparation; \citealt{Sousbie11}).

\HI\ in different environments has been studied at relatively high redshift by the Blind Ultra Deep \hi\ Environmental Survey (BUDHIES; \citealt{Verheijen07}) which observed \hi\ between $0.16<z<0.22$ around two clusters with the Westerbork Synthesis Radio Telescope.  The \hi\ detections trace infall linked to the larger filamentary structure and groups \citep{Jaffe12}.  Phase space diagrams with optical photometry permit galaxies to be recognized as having been stripped, or in various states of transformation \citep{Jaffe16}.  However, \hi\ morphology studies were not possible since all but the largest \hi\ detections are unresolved.

The COSMOS \HI\ Large Extragalactic Survey (CHILES)\footnote{\url{http://chiles.astro.columbia.edu}} is an ongoing \HI\ survey with the upgraded Karl G.~Jansky Very Large Array which is imaging, for the first time, \HI\ over a contiguous redshift range of $0<z<0.45$ \citep{Fernandez16}.  We have chosen to observe a single pointing in the COSMOS field for the extensive ancillary data that is available and the lack of bright radio continuum sources.  The CHILES pointing intersects large-scale galaxy over-densities at several redshifts: so-called ``walls'' at $z=0.12$, $z=0.19$, $z=0.37$, as well as the voids between them, and filaments which lace the entire volume.  
We have previously reported the detection of the highest redshift galaxy to date in \HI\ emission, an extremely atomic and molecular gas rich spiral at $z=0.376$ \citep{Fernandez16}.  
Ultimately, we expect the full survey to directly image over 300 galaxies in the COSMOS field, based on the scaling relations of \citet{Catinella12} and estimates from the HIMF.

In this paper we present results from the first 178 hours of observations of CHILES (referred to as `Epoch 1'), focusing on two redshift ranges coinciding with over-densities at $z\sim0.12$ and $z\sim0.17$.  One of the greatest challenges facing deep, high redshift \hi\ surveys is the increasing presence of interfering sources outside the protected \hi\ radio frequencies.  In this paper we demonstrate that with careful flagging and data calibration, we are still able to extract \hi\ spectral line sources from regions which are among the worst affected by RFI.
We discuss the lessons we have learned from training and implementing the automated \HI\ source finder, SoFiA \citep{Serra15}, on our large data cubes which will be directly applicable to future \HI\ surveys.  We also define the algorithm we use to create significance maps, essentially pixel-by-pixel signal-to-noise maps, and emphasize their importance in evaluating the reliability of the \hi\ morphology recovered by the source finder.  We characterize the spatial distribution of \HI\ in the galaxies in relation to their environment, which we measure in two ways: locally using the zCOSMOS 20k group catalogue, and globally using the DisPerSE algorithm to calculate the network of filaments in the COSMOS volume.  Our analysis is made possible through the extensive ancillary data available in the COSMOS field.

The paper is laid out as follows.  In Section \ref{data} we present the \HI\ data, optical spectroscopic redshift catalogue, and source finding strategy.  In Section \ref{environment} we describe two measures we use to quantify galaxy environment.  In Section \ref{results} we present the results of our source finding including resolved \HI\ total intensity and significance maps, and a summary of the gaseous and stellar properties of the galaxies.  In Section \ref{discussion} we discuss the properties of our detections in the context of their environment: both the local group or non-group environment to which they belong, as well as their position within the filamentary structure quantified by the distance to the nearest filament.  Finally, in Section \ref{summary} we summarize these results from CHILES and what we have learned for future \HI\ surveys.  Throughout this paper, we assume $H_0=70$~\kms~Mpc$^{-1}$, $\Omega_M=0.27$, and $\Omega_{\Lambda}=0.73$.

\section{Data \& Source Finding}
\label{data}

\subsection{\HI\ Spectral Line Data}
\label{speclinedata}

CHILES is a 1000 hour observing program with the NSF's Karl G. Jansky Very Large Array (VLA) seeking to detect the most \hi\ massive galaxies out to $z=0.45$.
Epoch 1 of CHILES consists of 178 hours of data taken in B-configuration between October 2013 and January 2014.  The CHILES pointing is centered at $10^h01^m24^s\, +02^{\circ}21^{\prime}00^{\prime\prime}$, and the observations use three frequency settings separated by 5 MHz to fill gaps in the frequency coverage created by recording the data over two subbands and 15 spectral windows.

The data is reduced in CASA 4.2 \citep{McMullin07} using version 1.2.0 of the NRAO VLA continuum pipeline\footnote{\url{https://science.nrao.edu/facilities/vla/} \url{data-processing/pipeline}} which we modified and tailored for the CHILES spectral line data.  We did extensive tests on how to flag the data in \texttt{RFLAG}, one of the automated flagging algorithms available in CASA.  The default setting is to allow the \texttt{RFLAG} algorithm to calculate statistics for each spectral window and then flag according to those values.  This is problematic for spectral windows severely affected by RFI, since the time and frequency rms will be high due to the presence of RFI, making it impossible to identify bad data. Instead we select a relatively clean spectral window near 1400 MHz to calculate the statistics, and scale that value for each spectral window with the expected variation in system temperature across the frequency range.  After the pipeline is run, we visually inspect the results for quality assurance, flag data if required, and rerun the pipeline as necessary.  \texttt{RFLAG} proved to be very effective and manual flagging was only necessary in redshift ranges severely affected by RFI.  This manual flagging was done primarily on the calibrators.  On the flux density calibrator, the automated flagging removed most of the high points, but left RFI wings in frequency which we flagged by-hand.  On the complex gain calibrator, we flagged bad time ranges or antennas.  Minimal flagging was done on the target and Figures \ref{fromnick1260} and \ref{fromnick1200} show that most of the RFI was successfully removed. 

The Epoch 1 image cubes were created using CASA and the facilities of Amazon Web Services (as described in \citealt{Dodson16}).  An initial model is subtracted from the visibilities with a spectral index for continuum sources both in the main lobe and outside the primary beam.  The model is channelised in 2 MHz frequency steps to account for the 17 MHz standing wave from the VLA, and binned into 0.5 hr time steps in hour angle to account for the movement of the sidelobes on the sky.  The data are imaged with a robustness of 0.8 in the CASA task \texttt{clean} in 4 MHz chunks of $64 \times 62.5$ kHz channels.  The cubes have a pixel scale of $2^{\prime\prime}$ and image $2.3\times2.3$ deg$^2$ to include strong continuum sources beyond the null of the primary beam.  We further clean the continuum emission in channels to remove the sidelobes of these strong sources.  Six 4 MHz chunks are then concatenated to produce several 24 MHz cubes with 4 MHz overlap between each.  The continuum subtraction is performed in the image plane with a first order fit.  When this was initially performed on the Epoch 1 data, the last 2 MHz were trimmed to save space, such that the final cubes are 22 MHz, corresponding to $352 \times 62.5$ kHz channels per cube.

\begin{figure}
\includegraphics[scale=0.45,clip,trim=28 0 60 460]{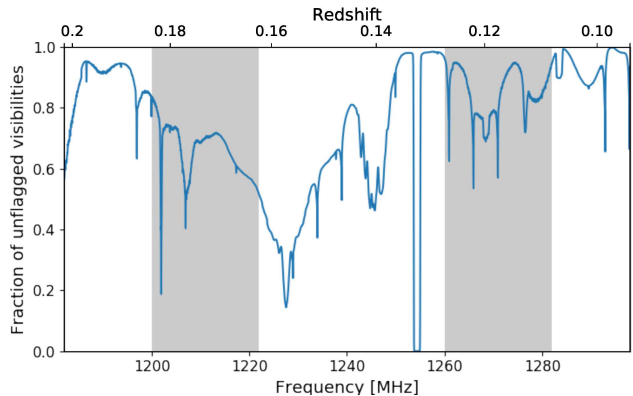}
\caption{The fraction of visibilities as a function of frequency remaining after flagging in the data reduction pipeline.  The grey areas represent the two frequency ranges discussed in this paper.  The blue line is the fraction of unflagged visibilites per channel in Epoch 1 of the CHILES survey.  The 1217-1237 MHz and 1240-1250 MHz ranges are affected by the GPS L2 and GLONASS L2 satellites, respectively, and are also seen in the pilot data (Figure 1 of \citealt{Fernandez13}). The 1268 MHz feature is due to the E6 transmision band on the COMPASS satellites which became operational after 2012.  The three narrow features that sit 5 MHz apart and cluster at each of 1200, 1235, 1270 MHz are due to the gaps between subbands that shift with our three different frequency settings.}
\label{rfi}
\end{figure}

\begin{figure*}
\includegraphics[scale=0.665,clip,trim=140 270 140 20]{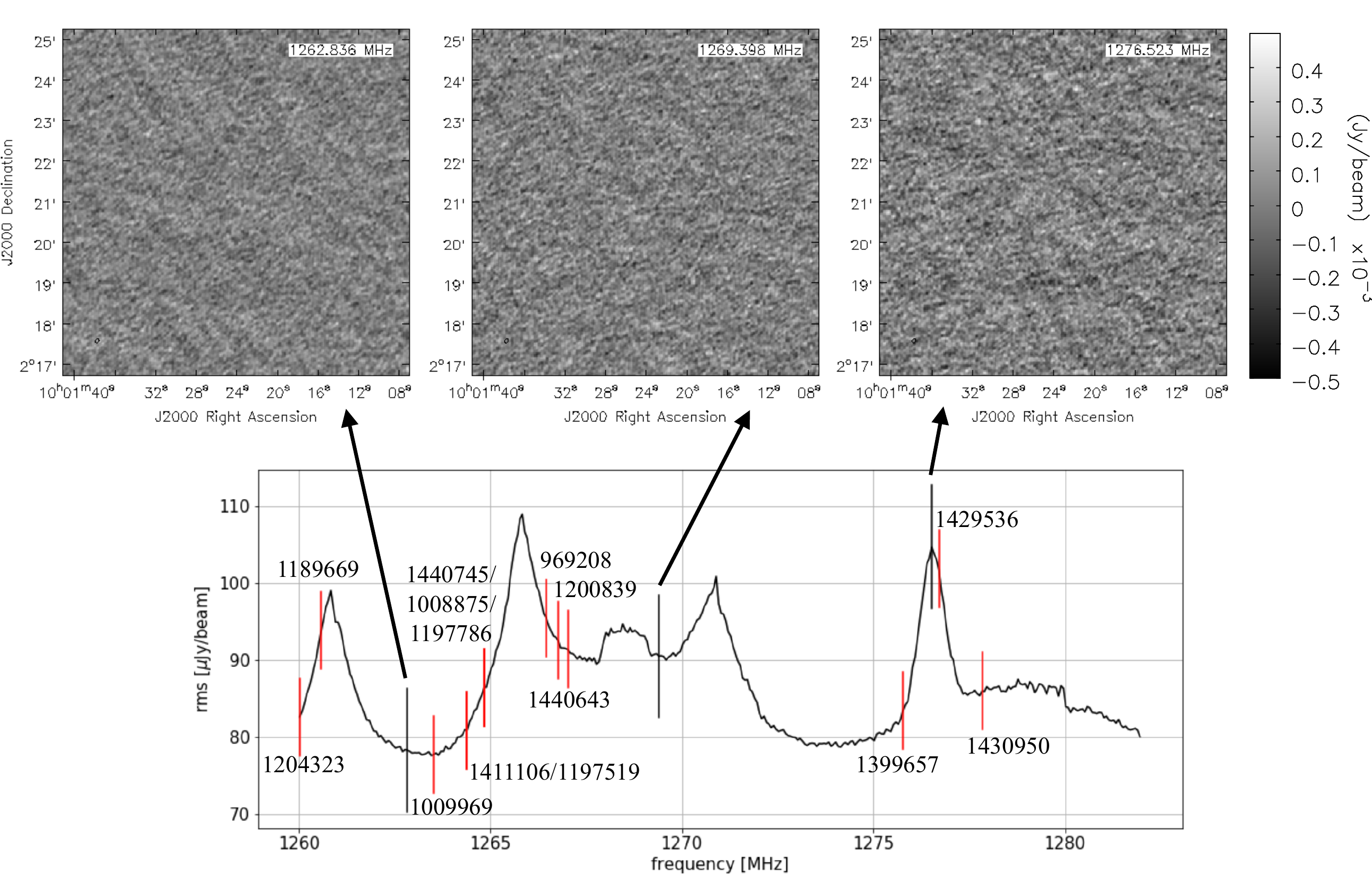}
\caption{The rms noise as a function of frequency covering the range of the z=0.12 wall, 1260--1282 MHz. Red lines indicate the placement of each detection along the spectral axis for the \hi\ spectroscopic redshift. The black lines indicate the emission-free channel in which the three sample channel maps are shown at low (left), intermediate (center), and high (right) relative rms.  The three peaks to the left are the gaps caused by the three different frequency settings.  The peak to the right, and other structure in the rms as a function of frequency reflect a larger percentage of data getting flagged as a result of RFI (see also Figure \ref{rfi}).  The lack of stripes in the channel maps demonstrate that we have successfully removed nearly all the RFI from these frequency ranges.}
\label{fromnick1260}
\end{figure*}

\begin{figure*}
\includegraphics[scale=0.665,clip,trim=160 270 180 20]{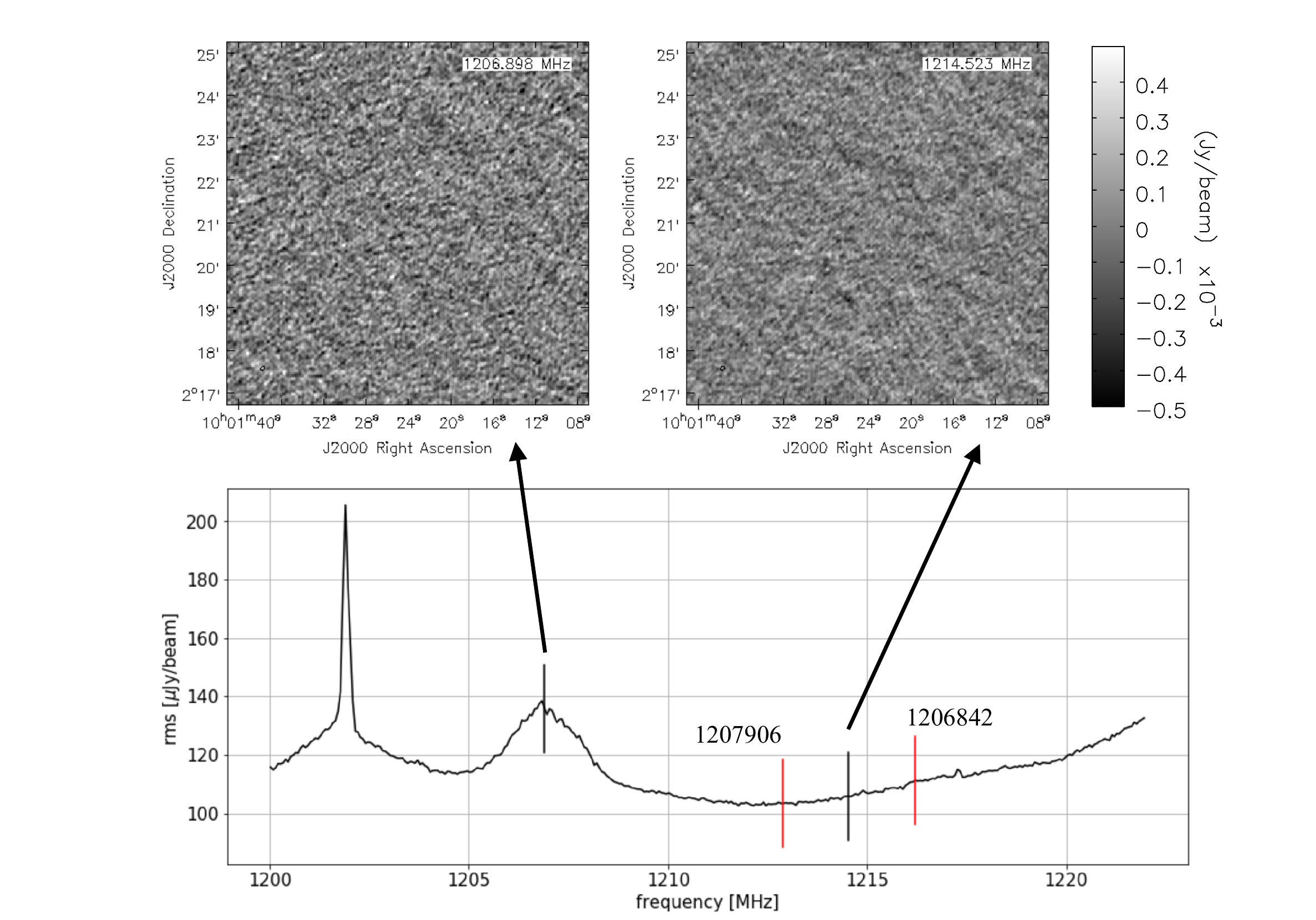}
\caption{The rms noise as a function of frequency covering the range of the z=0.17 volume, 1200--1222 MHz. Red lines indicate the placement of each detection along the spectral axis for the \hi\ spectroscopic redshift. The black lines indicate the emission-free channel in which the two sample channel maps are shown at high (left), and low (right) relative rms.  The sharp peak is caused by a gap due to the different frequency settings.  The broad peak is combination of the frequency gap and a larger percentage of data getting flagged as a result of RFI (see also Figure \ref{rfi}).  The lack of stripes in the channel maps demonstrate that we have successfully removed nearly all the RFI from these frequency ranges.}
\label{fromnick1200}
\end{figure*}

In this work, we focus exclusively on two frequency ranges: (1) 1260--1282 MHz which contain a redshift range of $0.108\le z\le0.127$ and to which we refer as the $z=0.12$ ``wall''; and (2) 1200--1222 MHz which corresponds to $0.162\le z\le0.183$ and which we call the $z=0.17$ ``volume''.  We selected the frequency range around the $z=0.12$ wall because it includes the most nearby over-density of galaxies in CHILES, and we selected the $z=0.17$ volume to test our pipeline and extraction of spectral line sources from the frequency range which is most strongly impacted by RFI (Figure \ref{rfi}).  In total, these two volumes represents less than 10\% of the total frequency coverage of the CHILES survey, but show that the overall data quality is good even in regions where much of the data is flagged due to RFI.

Figures \ref{fromnick1260} and \ref{fromnick1200} show the rms as a function of frequency for our two redshift ranges, as well as sample channels from the continuum subtracted line cubes.  Overall, we achieved a mean rms of 0.087 mJy beam$^{-1}$ per 62.5 kHz channel in the 1260--1282 MHz cube with about 10\% variation across the frequency range; and 0.11 mJy beam$^{-1}$ in the 1200--1222 MHz cube, with up to 20\% higher noise in ranges broadly affected by RFI, and less than a factor of 2 higher rms in a few of the worst affected channels.  The channel resolution corresponds to 16.4~\kms\ and 18.2~\kms\ in the observers frame, and the final image resolution is approximately 6.9 $\times$ 5.2 arcseconds and 7.0$\times$ 5.4 arcseconds, in the $z=0.12$ and $z=0.17$ cubes, respectively.
For source finding, we focused on the inner $34^{\prime}\times34^{\prime}$, which covers roughly 97\% of the full width at half power of the VLA primary beam at 1282 MHz (91\% at 1200 MHz).  These cubes cover an approximate volume of 4.3~$\times$~4.3~$\times$~94~Mpc and 6.0~$\times$~6.0~$\times$~110~Mpc, respectively.

\subsection{G10/COSMOS Spectroscopic Catalogue}

The Cosmic Evolution Survey (COSMOS; \citealt{Scoville07a}) was designed to study the concurrent evolution of galaxies, their star formation history, and active galactic nuclei with the assembly of large scale structure.  The survey compiles multi-wavelength imaging and spectroscopy from X-ray to radio wavelengths over approximately 2 deg$^2$, and includes the largest contiguous area imaged by Hubble Space Telescope (\textit{HST}; \citealt{Scoville07b}).

For both our source finding and data analysis, we use the multi-band photometry and optical redshift information from the G10/COSMOS v05 catalogue\footnote{\url{http://cutout.icrar.org/G10/dataRelease.php}}. The catalogue contains photometric data, and spectroscopic measurements for galaxies in the COSMOS region obtained from both an independent re-analysis of the zCOSMOS spectra, and direct zCOSMOS, PRIMUS \citep{Coil11,Cool13}, VVDS \citep{Garilli08}, and SDSS \citep{Ahn14} redshifts in the region.  It also provides various combinations of these redshifts to assign quality flags to each redshift (see \citealt{Davies15} for details). Version 5 of this catalogue contains updated redshifts from the zCOSMOS DR3 data release, the COSMOS2015 photometric redshifts \citep{Laigle16}, and the redshift matching assignment in the quality flags has been updated.  Full details of the latest catalogue can be found in \citet{Andrews17}.

\subsection{\HI\ Source Finding}
\label{sourcefinding}

We conducted extensive inspection of the two data cubes by-eye, followed by semi-automated source finding.   We used the G10/COSMOS v05 source list as input for a systematic visual search of the cubes to build a list of `candidate detections' coincident with spectroscopic counterparts.  This included 211 (104) galaxies with spectroscopic redshifts in the $z=0.12$ wall ($z=0.17$ volume).  The first several passes were generous in our by-eye evaluation of what was considered a candidate.  We included objects that appeared continuous at the $3\sigma$ level in at least 2-3 channels, and sometimes at lower sigma if it was contiguous over many channels and appeared to form a spatially and spectrally coherent structure.  This was similar to the source finding approach taken with the CHILES pilot data \citep{Fernandez13}.  Spatial coherence between channels generally meant that the peak did not move by more than a beam width, but in some cases this criterion was relaxed to a couple beams if we were only picking up the two peaks of a double horned profile on opposite sides of a galaxy.  We then smoothed the data to twice the velocity resolution in both cubes from 16 (18)~\kms\ to 32 (36)~\kms\ and performed the search by-eye again.

In addition, we searched the cube by-eye for potential \HI\ sources that lacked known optical spectroscopic counterparts.  We included them in the testing described below if they had an photometric counterpart in \textit{HST} imaging.  In the end, however, there were no \hi\ candidates in this category that were confirmed as detections.

The generous list of candidates became a training set for SoFiA, the Source Finding Application \citep{Serra15}, where the candidates generally fell into three subjective categories: those confident detections which are easily identified; more marginal detections whose properties still strictly meet the criteria of being above the 3$\sigma$ level in several channels and appear at the same velocity as an optical redshift but are closer to the noise level; and those which do not strictly meet the criteria selected above, but are included because they are persistent when the channels are viewed quickly like frames of a movie.  In particular the last category is included in the case that through iterative smoothing in space and velocity, SoFiA may pick up detections that are missed by-eye.

This list of candidate detections was then run through the SoFiA source finder (version 1.1.0) using the targeted search option\footnote{The code is available at \url{https://github.com/SoFiA-Admin/SoFiA}} on noise flattened dirty images.  The targeted search extracts cubelets around specified positions to look for detections: the smaller the cubelet, the faster the source finder runs, but the larger the cubelet the better the noise is characterized, which is then used to reject false detections using a reliability function.  We chose a small subset of candidate sources--both strong and weak candidates, with broad and narrow \HI\ profiles, to tune the source finding parameters and cubelet sizes.

Ultimately we found that no single run of SoFiA would detect all the \hi\ sources.  First, we optimized two runs of `smooth and clip' with the source finder \citep{Popping12}.  In both cases we used a low detection threshold of $3\sigma$.  The volumes we searched were $54\times54$ arcseconds and 1.2 or 5 MHz in frequency, which correspond to $27\times27\times20$ and $27\times27\times80$ voxel cubes.  In physical units, these cubelets are approximately $110\pm10$ kpc $\times$ 315~\kms\ or 1312~\kms\ in the $z\sim0.12$ cube, and $155\pm10$ kpc $\times$ 348~\kms\ or 1452~\kms\ in the $z\sim0.17$ cube, centered on the optical spectroscopic detection.
In order to account for variable noise characteristics throughout the cube (Figures \ref{fromnick1260} and \ref{fromnick1200}), we allowed SoFiA to scale the noise by channel.

In addition to evaluating whether the source finder found a source at the expected location, we judged the performance of a set of parameters by the number of false detections that arose, as well as their shape--for example by excluding single pixel objects, or objects that were extended in the same directions as residual artefacts (faint stripes) in the data.
Merging parameters in the source finder helped to prevent large sources from being broken into several smaller ones by merging peaks found within a few pixels in right ascension, declination, and velocity.  We also specified minimum size thresholds to eliminate unrealistically small (below the size of the beam) sources.

Finally, we used the reliability parameters to reduce the number of false detections.  The reliability is determined by parameterizing both positive and negative detections, assuming all negative detections are noise, and then determining a probability that a positive detection is real.  Sources below a probability of 95\% were rejected.  The method works best with a low signal-to-noise threshold so that a substantial number of negative noise peaks are detected and the reliability can be determined in a statistically meaningful way.  The method is explained in \citet{Serra12}.  Additionally, we set a minimum total flux value for reliable positive detections.  
These runs of smooth-and-clip were the most restrictive in that they only found the brightest candidates, but we attempted to optimize the parameters to find as many by-eye candidate detections as possible, excluding as many false-positives as possible without sacrificing detections.  We compared the masks and moment maps output by SoFiA with the original data cubes and with overlays on optical images to judge the source finder's performance.  \HI\ detections had to be spatially coincident with the optical galaxy and within $\sim$200~\kms.  Ultimately 10.5 out of 16 of our reported detections were parameterized in these runs of SoFiA (see below for an explanation of the 'half detection').

The third run on SoFiA turned off the reliability function, but used pixel dilation in order to `force' the program to find more marginal detections.  Pixel dilation is a method of optimally growing the mask around detections and it is described in detail in \citet{Serra15}.  We experimented with different dilation parameters.  Sources in the cube tend to be small (only a few beams across), so we set a maximum dilation of 3 pixels in order to avoid merging noise peaks due to our low detection threshold.  Since the reliability function was turned off, we also increased the signal-to-noise cutoff to $3.5\sigma$.  This run of SoFiA found 4.5 more \HI\ detections.

As mentioned above, in the case of J100136.9+023032.1, one half of the double peaked \hi\ profile was found using the first run of SoFiA with the reliability parameters.  Unfortunately, even with including larger smoothing kernels up to 15 channels, we could not train SoFiA to find both peaks simultaneously.  We suspect this is because there were 8 channels between the two halves without detected \HI\ emission, and smoothing the data did not add significant signal compared to the noise to link the two detections.  The second half of the galaxy could only be detected and characterized using the pixel dilation method.  In the results presented in Section \ref{results}, we have added the masked emission from both halves of the galaxy to create the total intensity map, but we have integrated the \HI\ profile over the entire width of the galaxy to estimate the total \HI\ mass.  This broad line source is difficult to pick out even in the integrated spectrum.  It is an example in which human interaction with the source finder is important to recover the entire galaxy.

The results of the optimization were that SoFiA did not find all of the candidates that we generously (and subjectively) included from our by-eye inspection, and in fact we only report approximately 40\% of our original candidates from these frequency ranges that were confirmed with SoFiA. (The original list of candidates included approximately 35 in the $z=0.12$ cube and 5 in $z=0.17$ cube.)  
The conclusion from our testing is that after optimization, SoFiA can provide a sanity check and a reasonable objective cutoff for \hi\ detections.  The candidates will be confirmed or refuted with deeper CHILES imaging to come.  

After the final \hi\ detections were identified, we cleaned the \hi\ spectral line emission in CASA using the masks output from SoFiA.  The emission within the mask was cleaned to a threshold of 1$\sigma$ of the rms.  We then ran the cleaned cubes through SoFiA again to produce total intensity maps and other final data products which are presented in Section \ref{results}.

Of course, SoFiA can also be operated in a blind mode (e.g.~\citealt{Westmeier17}).  The comparison between the blind and targeted modes run on the CHILES data will be the subject of a future paper.  Here we simply note that in addition to the much larger time required to run the blind mode, out of necessity, this mode requires a higher noise threshold, and would likely miss the faintest sources.  The 22 MHz cubes covering the primary beam consist of 369 million voxels.  In an optimistic case of perfectly Gaussian noise, one would still expect $\sim$26 independent voxels to be above $+5\sigma$ (and $\sim$2.9$\times10^3$ voxels above $+4\sigma$).  
Reliability parameters, and the criterion that detections be more than one contiguous voxel narrow the number of candidate detections, but nonetheless, a fully automated source finder still requires quite a bit of human interaction for interpretation with the quality of the current data.

To summarize: we did not conduct a blind \hi\ search, but used the COSMOS/G10 v05 spectroscopic redshift catalogue to run a targeted search for \hi\ objects associated with known optical counterparts.  All 16 \hi\ sources were found by SoFiA, but it took three different sets of parameters.  All sets of parameters used SoFiA's ``smooth and clip'' method of source finding.  Two sets of parameters had the reliability function turned on; the third set of parameters used pixel dilation with the reliability function turned off.  Despite using a very low signal-to-noise cutoff of 3 to 3.5, SoFiA only detected 40\% of the original candidate training set.  The remaining 60\% were rejected because they did not hit the minimum signal-to-noise cutoff, were determined to have low reliability, or were smaller than the size of the CHILES beam.

\subsection{The challenge of radio frequency interference}

One of the main challenges for the CHILES survey is RFI from satellites as well as ground based radar.  It is an unfortunate fact that our strongest wall is in one of the worst RFI plagued frequency ranges.  
In general, the frequency range from 1200 MHz to 1282 MHz is most affected by RFI radar at the Albuquerque airport,  the GPS L2 and Glonass L2 satellites at 1217-1237 and 1240-1250 MHz\footnote{\url{https://science.nrao.edu/facilities/vla/observing/RFI/} \url{L-Band}}, as well as the COMPASS satellites around 1268 MHz which became operational in 2012\footnote{\url{https://en.wikipedia.org/wiki/BeiDou\_Navigation\_} \url{Satellite\_System\#BeiDou-2} and references therein.}.  As mentioned in Section \ref{speclinedata}, the vast majority of RFI was identified by our automated flagging strategy with a very small fraction needing to be flagged by-hand on the target source.  Most of the manual flagging was conducted on the calibrators to ensure good calibration solutions.

Figure \ref{rfi} shows the fraction of target visibilities remaining after flagging in the data reduction pipeline across the range from 1150 to 1300 MHz in our final data cube.  The shaded areas show the frequency range under consideration here.  As can be seen, they fall just outside the most RFI degraded data ranges.  We show in Figures \ref{fromnick1260} and \ref{fromnick1200} the rms as a function of frequency and grey scale representations of some of our channels.  The peaks in rms correspond closely to the dips in the number of visibilities at those frequencies in Figure \ref{rfi}.  Nonetheless, the channel maps in Figures \ref{fromnick1260} and \ref{fromnick1200} show there are no significant stripes remaining due to RFI.  Thus we conclude that even in the frequency ranges most seriously affected by RFI after careful flagging of the data we are able to recover \HI\ sources in these ranges of the spectrum at a high confidence.

\section{Characterizing Environment}
\label{environment}

Numerous studies have sought to characterize environments in the COSMOS volume.  They all rely either primarily or secondarily on redshift information from optical detections, whether they use neighbors to associate galaxy groups \citep{Knobel12}, infer galaxy clusters from diffuse X-ray emission \citep{George11}, or map the connections of filaments and voids \citep{Cybulski14,Darvish17}.

In this work, we use the existing zCOSMOS 20k galaxy group catalogue from \citet{Knobel12} to cross match the counterparts of our \HI\ detections with known galaxy group members and their group properties.  In this study, we consider whether a galaxy belongs to a group or not, and the properties of the group to which it may belong, a measure of the \textit{local} environment.  The groups are presumed to be gravitationally bound, where the member galaxies reside within a common parent dark matter halo.

In addition, we use the DisPerSE algorithm (Discrete Persistent Structures Extractor; \citealt{Sousbie11}) to locate galaxies in filaments, and voids.  We consider whether a galaxy lives on a filament or not, and its distance to the nearest filament as a measure of its \textit{large-scale} (or global) environment. Filaments are not gravitationally bound, but presumed to be higher density `ridges' on to which material flows from voids, and which give the cosmic web its structure.  Filamentary structure is generally larger in scale and lower in density in terms of galaxies per unit volume than individual groups.  Galaxy clusters and large groups tend to reside at the intersection of filaments \citep{Bond96}.

\subsection{zCOSMOS 20k Group Catalogue}

\begin{figure*}
\includemedia[3Dtoolbar,3Dmenu,3Droll=0.13605703476132344,
3Dc2c=-0.6208851933479309 0.7399421334266663 0.258819043636322,
3Dcoo=0 -168829.4375 -176.673828125,
3Droo=209776.68683215926,
3Dlights=Headlamp,
transparent=false,width=1.0\textwidth, height=1.0\textwidth,activate=onclick]{\includegraphics[scale=0.5,clip,trim=0 100 0 100]{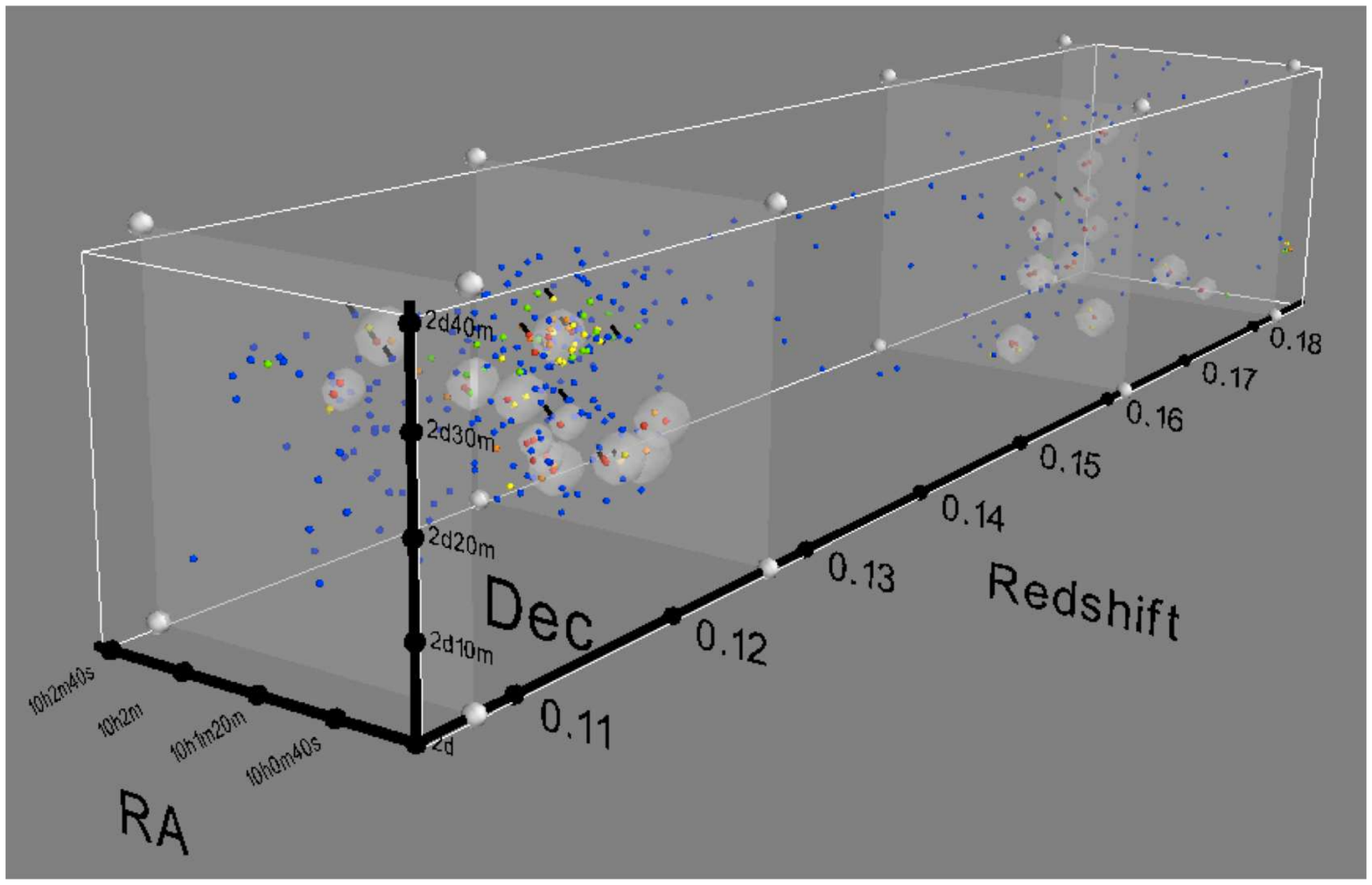}}{cone_v2.u3d}
\caption[]{Interactive 3D figure 
(must be opened with Adobe Acrobat) showing the galaxy distribution in the two redshift ranges, including all galaxies with spectroscopic redshifts in the G10/COSMOS v5 catalogue \citep{Davies15,Andrews17}.  Large transparent spheres indicate the position of zCOSMOS 20k galaxy groups \citep{Knobel12}.  Blue symbols represent non-group members or galaxies with a probability of belonging to a group of less than 50\%. Green to red symbols have 50--100\% likelihood of belonging to a group.  Galaxies which we have detected in \HI\ are indicated by a black diagonal line pointing at their symbol.  The transparent planes represent the bounds of the two redshift volumes around $z=0.12$ and $z=0.17$. The redshift scale has been compressed by a factor of 10 relative to the spatial axes. The field of view at a fixed opening angle is a cone in real space rather than a rectangular prism, so the RA/Dec tickmarks have been calculated relative to the center of the frequency range at z=0.14.}
\label{3d}
\end{figure*}

The zCOSMOS 20k group catalogue \citep{Knobel12} used a friends-of-friends (FoF) group finding algorithm on zCOSMOS galaxies with `high-quality' spectroscopic redshifts.  The spectroscopic selection included galaxies with $I_{\text{AB}}\le22.5$ in the redshift range $0.1\le z\le1.2$.  The group finding algorithm used successive re-runs of varied linking length parameters to identify the richest groups first, down to the lowest mass groups, while trying to avoid contamination and overcome fragmentation.  Galaxies with photometric redshifts were then associated with groups using a probabilistic approach to improve the overall group statistics.  

The overall spectroscopic completeness of zCOSMOS 20k, on which the group catalogue is based, is estimated to be 50\%, although completeness varies across the COSMOS survey area.  However, if one includes the photometric sources, it is estimated that the group membership completeness improves to between 75\% and 95\% for groups of all masses.  At $z=0.15$, the FoF algorithm is estimated to detect 95\% of galaxy groups with dark matter halos more massive than $10^{12.5}$ \msun, but falls to 65\% for groups with halos greater than $10^{12.0}$ \msun\ (Figure 2 of \citealt{Knobel12}) indicating that the catalogue is not good at recovering low mass groups.  

In the group catalogue, galaxies are occasionally matched with multiple groups, especially if they only have photometric redshifts.  In this case, we assume the galaxy belongs to the group with the highest probability association (PA) match (Section 4.2 of \citealt{Knobel12}).  Further, for our analysis we only consider galaxies as group members if they have a PA greater than 0.5.  Galaxies with spectroscopic redshifts generally have a PA greater than 0.7.

One caveat of using the group catalogue for environment, and the G10/COSMOS v5 catalogue as a basis for a targeted \HI\ search is that reprocessing of the zCOSMOS data has increased the number of objects for which there are reliable redshifts.  These spectroscopic redshifts were not included in the zCOSMOS 20k data set on which the FoF algorithm was run, but their group membership was estimated from photometric redshifts and overall group properties.  In Section \ref{discussion}, we discuss galaxy group candidates with photometric redshifts in the 20k data that now have spectroscopic and \HI\ redshifts.  We also look at individual galaxies that were not assigned to a group, but which appear to have disturbed \HI\ and a nearby companion just outside the FoF linking lengths.

Figure \ref{3d} shows an interactive three dimensional representation of galaxies with G10/COSMOS spectroscopic redshifts, and zCOSMOS 20k galaxy groups which include both the $z=0.12$ wall and $z=0.17$ volume.  The small coloured symbols represent galaxies where colours green to red have 50-100\% probability of belonging to a galaxy group, and blue symbols have less than 50\% probability of being a group member.  The large transparent spheres represent the geometric center of galaxy groups, and the frequency coverage of our two data cubes are bounded by transparent planes.  It is important to note that the redshift axis is 45 times more compact than the spatial axes in this figure illustrating, as has been seen in other group catalogues, velocity dispersion is the most difficult parameter to accurately recover in group catalogues and leads to both interlopers and incompleteness \citep{Berlind06,Duarte14}

In total, there are 11 groups in the $z=0.12$ wall and 11 groups in the $z=0.17$ volume probed here.  These are typically low mass groups and nearly all have only 2-3 spectroscopic members.  The most massive group in the Knobel catalogue within the CHILES volume has 30 spectroscopic members and a dark matter halo mass of $10^{13.6}$~\msun, and falls in the $z=0.12$ wall, making it the most massive gravitationally bound structure in the entire CHILES (Table 2).  The second most massive group in the same redshift range has 4 spectroscopic members and a halo mass of $10^{12.9}$~\msun; the rest are below $10^{12.6}$~\msun.  The 11 groups at $z=0.17$ are all estimated to have a halo mass less than $10^{12.2}$~\msun.

\subsection{DisPerSE}
\label{dispersedesc}

The Discrete Persistent Source Extractor, DisPerSE, is a scale free topological algorithm that identifies various types of structures in a given distribution of particles \citep{Sousbie11}. In particular, the algorithm was developed with the intention of defining the structures of the cosmic web both in large redshift surveys (e.g.~\citealt{Kleiner17}) and N-body simulations (e.g.~\citealt{Libeskind17}). The method is run on a density field of the particles where topological features corresponding to differently indexed manifolds, in terms of cosmological environments, voids, walls, filaments and clusters, are calculated and returned.

For ease of interpretation and to take advantage of the planar structure of the $z=0.12$ wall, we run DisPerSE in its two dimensional form.  We explored different parameters for the DisPerSE functions and the best combination of boundary type and significance level, considering the small field of view, and then applied them to redshift slices in the CHILES range compressed along the redshift axis.  We run DisPerSE over the full 2$\times$2 deg$^2$ COSMOS field of view to ensure the reliability of structure on the edge of the roughly 0.5$\times$0.5 deg$^2$ CHILES field.  The chosen DisPerSE parameters are a periodic boundary condition with \textit{nsig} $=2$, using only high quality spectroscopic redshifts.  The full exploration of the DisPerSE parameters with the G10/COSMOS data will be described in a forthcoming paper by N.~Luber et al.  

In this application, the width of the redshifts slices before compression is $\Delta z=0.01$, however, we only calculate the distance from a galaxy to the cosmic web if this galaxy is in the central $\Delta z=0.005$ for that slice.  The ``critical points'' are the tracers of the filamentary network of the cosmic web returned by DisPerSE.  We calculate the projected distance of a galaxy to the cosmic web by finding the distance of that galaxy from the nearest critical point, and then dividing by the average separation of critical points in that redshift bin.  The position of galaxies in relation to the filamentary structure is discussed in Section \ref{filamentvoid}.

\begin{table*}
\centering
\caption{\HI\ detections and their properties.}
\begin{tabular}{cccccccccc}  
\hline
 COSMOS & GAMA  & Optical Position & \HI\ (Optical) & \HI\ Mass     & \HI\ $W_{\text{int}}$  & Stellar Mass      & SFR                   & Group &        \\    
 ID\_08 & v5 ID & (J2000)          & Redshift   & $\log(M_{\odot})^a$ & (km s$^{-1}$)$^b$    & $\log(M_{\odot})^c$ & $M_{\odot}$~yr$^{-1}$ & ID$^d$ & Figure \\     
 (1)    & (2)   & (3)             & (4)       & (5)               & (6)           & (7)               & (8)                   & (9)    & (10) \\     
\hline
1008875 & 6008391 & 10$^h$00$^m$51.1$^s$ +2$^{\circ}$11$^{\prime}$37.7$^{\prime\prime}$ & 0.1227 (0.1226) & 9.40 &  66 & 9.60\phantom{*} & 0.33\phantom{*} & 300 & \ref{mom0c} \\  
1009969 & 6005372 & 10$^h$00$^m$53.1$^s$ +2$^{\circ}$10$^{\prime}$57.7$^{\prime\prime}$ & 0.1260 (0.1238) & 9.59 &  83 & 9.56\phantom{*} & 0.62\phantom{*} & (300) & \ref{mom0c} \\  
1430950 & 6002394 & 10$^h$01$^m$11.6$^s$ +2$^{\circ}$32$^{\prime}$27.1$^{\prime\prime}$ & 0.1115 (0.1116) & 9.66 & 212 & 8.92\phantom{*} & 0.51\phantom{*} &  -- & \ref{mom0b}a \\  
 969208 & 6008198 & 10$^h$01$^m$16.7$^s$ +2$^{\circ}$17$^{\prime}$11.9$^{\prime\prime}$ & 0.1215 (0.1216) & 9.87 & 547 & 10.5\phantom{*} & 5.87\phantom{*} &  -- & \ref{mom0e} \\  
1440745 & 6006194 & 10$^h$01$^m$19.8$^s$ +2$^{\circ}$27$^{\prime}$56.5$^{\prime\prime}$ & 0.1231 (0.1230) & 9.67 &  66 & 8.79\phantom{*} & 0.24$^f$        &   1 & \ref{mom0a}a \\  
1206842 & 6020163 & 10$^h$01$^m$19.9$^s$ +2$^{\circ}$23$^{\prime}$10.1$^{\prime\prime}$ & 0.1677 (0.1679) & 9.94 & 180 & 10.3\phantom{*} & 3.91\phantom{*} & (618) & \ref{mom0d} \\  
1197786 & 6001872 & 10$^h$01$^m$20.3$^s$ +2$^{\circ}$27$^{\prime}$14.9$^{\prime\prime}$ & 0.1234 (0.1233) & 9.70 & 133 & 9.45\phantom{*} & 0.57\phantom{*} &   1 & \ref{mom0a}b \\  
1197519 & 6008761 & 10$^h$01$^m$20.5$^s$ +2$^{\circ}$18$^{\prime}$17.8$^{\prime\prime}$ & 0.1229 (0.1230) & 9.83 & 365 & 10.3\phantom{*} & 7.20\phantom{*} &  -- & \ref{mom0e} \\  
1207906 & 6005750 & 10$^h$01$^m$21.9$^s$ +2$^{\circ}$22$^{\prime}$46.8$^{\prime\prime}$ & 0.1706 (0.1707) & 9.80 &  72 & 8.27\phantom{*} & 0.31$^f$        & 618 & \ref{mom0d} \\  
1204323 & 6005730 & 10$^h$01$^m$23.2$^s$ +2$^{\circ}$24$^{\prime}$15.0$^{\prime\prime}$ & 0.1269 (0.1269) & 9.49 &  50 & 10.1\phantom{*} & 3.89\phantom{*} &  -- & \ref{mom0b}b \\  
1440643 & 6022540 & 10$^h$01$^m$26.9$^s$ +2$^{\circ}$28$^{\prime}$02.9$^{\prime\prime}$ & 0.1219 (0.1211) & 9.69 & 116 & 8.48$^e$        & 0.59\phantom{*} &   1 & \ref{mom0a}c \\  
1200839 & 6020082 & 10$^h$01$^m$31.5$^s$ +2$^{\circ}$25$^{\prime}$57.7$^{\prime\prime}$ & 0.1213 (0.1250) & 9.65 & 100 & 9.35$^g$        & 1.28$^{fg}$     &   1 & \ref{mom0a}d \\  
1429536 & 6002388 & 10$^h$01$^m$33.8$^s$ +2$^{\circ}$33$^{\prime}$03.0$^{\prime\prime}$ & 0.1127 (0.1126) & 9.65 & 195 & 9.39\phantom{*} & 0.56\phantom{*} & 296 & \ref{mom0a}f \\  
1411106 & 6013987 & 10$^h$01$^m$36.9$^s$ +2$^{\circ}$30$^{\prime}$32.1$^{\prime\prime}$ & 0.1236 (0.1238) & 9.92 & 566 & 10.5\phantom{*} & 5.87\phantom{*} &   1 & \ref{mom0a}e \\  
1399657 & 6002302 & 10$^h$01$^m$57.4$^s$ +2$^{\circ}$35$^{\prime}$08.8$^{\prime\prime}$ & 0.1133 (0.1134) & 9.84 & 212 & 9.54\phantom{*} & 0.32\phantom{*} &  -- & \ref{mom0b}c \\  
1189669 & 6001862 & 10$^h$02$^m$09.4$^s$ +2$^{\circ}$20$^{\prime}$52.1$^{\prime\prime}$ & 0.1266 (0.1268) & 10.1 & 402 & 10.2\phantom{*} & 6.14\phantom{*} &  -- & \ref{mom0b}d \\  
\hline
\end{tabular} \\
\label{properties}
$^a$Uncertainties on the \hi\ mass are of order 20\%. For barely resolved sources, it may be of order 50\%.  See text for details.
$^b$\hi\ widths are optical velocity in the observer's frame and are simply the width over which we intergrate the \hi\ profile (see Figures \ref{mom0a}--\ref{mom0e}).
$^c$Uncertainties on the stellar mass are of order 25\% and uncertainties on the star formation rate are of order 35\%.  See text for details.
$^d$Galaxies in parentheses have \hi\ redshifts which place them outside the Knobel determined group.
$^e$No counterpart in the Spitzer IRAC bands so stellar masses are calculated using the absolute magnitudes estimated by SED fits. 
$^f$UV+IR star formation rate cannot be calculated because there are no infrared counterparts so the SFRs are quoted from the COSMOS 30-band photometric redshifts catalogue \citet{Ilbert09}.
$^g$Source is broken into two objects in the COSMOS 20k and G10/COSMOS catalogues.  The second source has no spectroscopic redshift but has the following COSMOS and GAMA IDs, respectively: 1200838, 6275580 (see row (d) in Figure \ref{mom0a}).  To estimate the stellar mass and star formation rates, we have added the values calculated independently for each source.  The table columns are described in Section \ref{props}.
\end{table*}


\section{Results}
\label{results}

With the data and methods described throughout Section \ref{data}, we detect 16 \HI\ sources in Epoch 1 of the CHILES survey (Table \ref{properties}).  Of these, 14 sources reside in the data cube spanning 1260--1282 MHz ($0.108\le z\le0.127$), and 2 sources in the cube spanning 1200--1222 MHz ($0.162\le z\le0.183$).  Taking $\sigma$ as the mean rms noise in the cubes, the 5$\sigma$ \HI\ mass sensitivity of our observations over 100~\kms\ is $10^{9.3}$~\msun\ at $z\sim0.12$, and $10^{9.8}$~\msun\ at $z\sim0.17$.  Thus, we are sensitive to \HI\ objects below the knee of the $z=0$ \HI\ mass function in the $z=0.12$ wall, and essentially at and above the knee in the $z=0.17$ volume \citep{Martin10,Jones18}.  It is important to remember that the volume searched represents less than 10\% of the total frequency coverage of the survey.

\begin{figure*}
\includegraphics[]{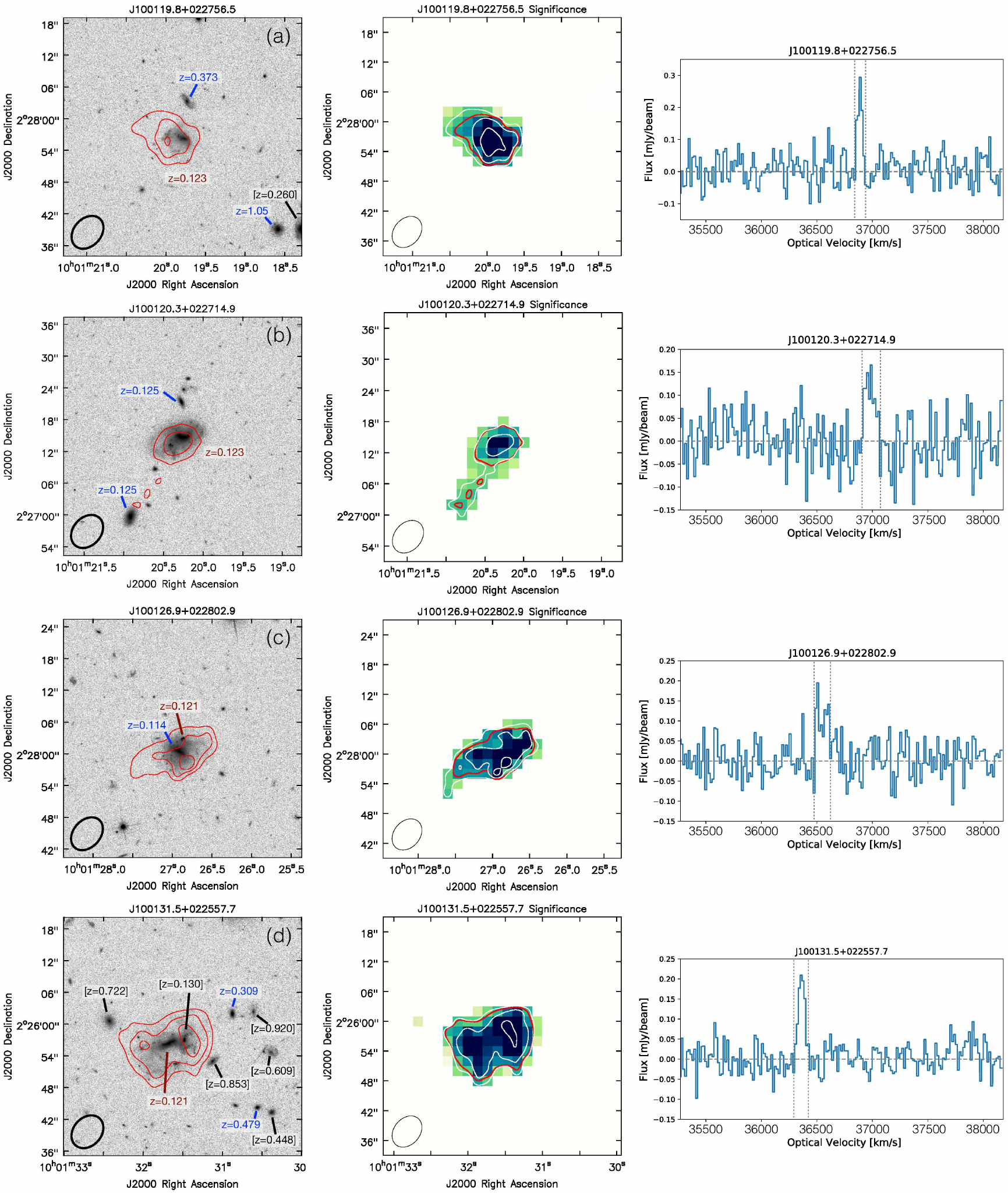}
\caption{Group galaxies in the $z=0.12$ wall. Left: \HI\ contours overlaid on $45^{\prime\prime}\times45^{\prime\prime}$ \textit{HST}/ACS I-band (F814W) images.  Contours are (1.6, 2.7, 5.3, 8, 13, 21)$\times10^{20}$ cm$^{-2}$.  Galaxies with spectroscopic redshifts are in blue (or red if they belong to groups).  Bright galaxies with photometric redshifts are in black and bracketed.  Center: Pixel-by-pixel signal-to-noise ratio. Contours are $\Sigma=(2,4,6,8)$, and the color scale is saturated at $\Sigma=5$.  We have included the lowest \hi\ column density contour from the total intensity map in red. Right: Integrated \hi\ line profile, enclosing lowest contour in left image.  The dashed vertical lines indicate the width over which the emission is integrated to calculate the \hi\ mass in Table \ref{properties}. Rows (a)-(d): these galaxies with red contours belong to Knobel Group 1, the most massive group in the CHILES volume.}
\label{mom0a}
\end{figure*}

In Figures \ref{mom0a}--\ref{mom0e} we present the \HI\ total intensity maps of each of our detections overlaid as contours on \textit{HST} Advanced Camera for Surveys (ACS) I-band (F814W) mosaic images obtained from the COSMOS Archive and IRSA cutout service\footnote{\url{http://irsa.ipac.caltech.edu/data/COSMOS/index_cutouts.html}} \citep{Koekemoer07,Massey10}.  We inspected the $90^{\prime\prime}\times90^{\prime\prime}$ area around each detection for galaxies with nearby spectroscopic or photometric redshifts.  Image sizes are chosen to show nearby galaxies which may be interacting with \hi\ detections, or to include nearby optical group members when possible.  The \HI\ total intensity maps are generated using the masks from the automated SoFiA source finding, and the contours correspond to 3, 5, 10, 15, 25, 40 times the mean rms in a single channel.  

The reliability of the \hi\ morphology in the total intensity maps is a key question in this study. It is made more complicated because the number of channels contributing to the total \HI\ flux varies per pixel.  Therefore, in Figures \ref{mom0a}--\ref{mom0e} we also present maps of the pixel-by-pixel signal-to-noise ratio, which we call the ``significance'' of the \HI\ emission.  These are calculated using the following equation:
\begin{equation}
\Sigma(i,j) = \frac{I(i,j)}{\sigma \sqrt{N_{\text{chan}}(i,j)}},
\end{equation}
where $\Sigma$ is the significance in the $(i,j)$th pixel, $I(i,j)$ is the total \hi\ intensity in said pixel, $\sigma$ is the mean rms per channel over the entire cube, and $N_{\text{chan}}(i,j)$ is the number of channels contributing to the emission in that pixel.  For example, if a pixel has an intensity equal to the mean rms in a single channel, then the significance, $\Sigma=1$.  In general, the strong detections in our study reach peak significance values greater than 6.  Detections with significance contours less than or equal to 4, should be treated with more caution.  The full CHILES survey will provide an important check on the ultimate reliability of the low signal-to-noise emission.

The \HI\ detected galaxies span a range of local environments: five out of 16 live in a massive group with at least 30 members; three appear to live in low mass groups with only 2-3 close members; three galaxies have nearby companions with which they may not be gravitationally bound, but may have had a recent interaction; two are former group members that with new \HI\ redshifts appear be single galaxies located on the backbone of a filament; and three do not belong to a known group in the Knobel catalogue and do not have obvious nearby companions.  

In terms of the large-scale environment, 10 out of 16 of the \HI\ detections live within 100 kpc of a filament; while the remaining six live at distances greater than 180 kpc from a filament.  The most isolated galaxy in our sample lives $\sim$250 kpc from the nearest filament, although this is still not among the most isolated galaxies for which we have optical spectroscopic redshifts.  In Section \ref{discussion} we discuss group membership and environment in detail, especially in relation to the overall \HI\ properties.

\renewcommand{\thefigure}{\arabic{figure} (Cont.)}
\addtocounter{figure}{-1}
\begin{figure*}
\includegraphics[]{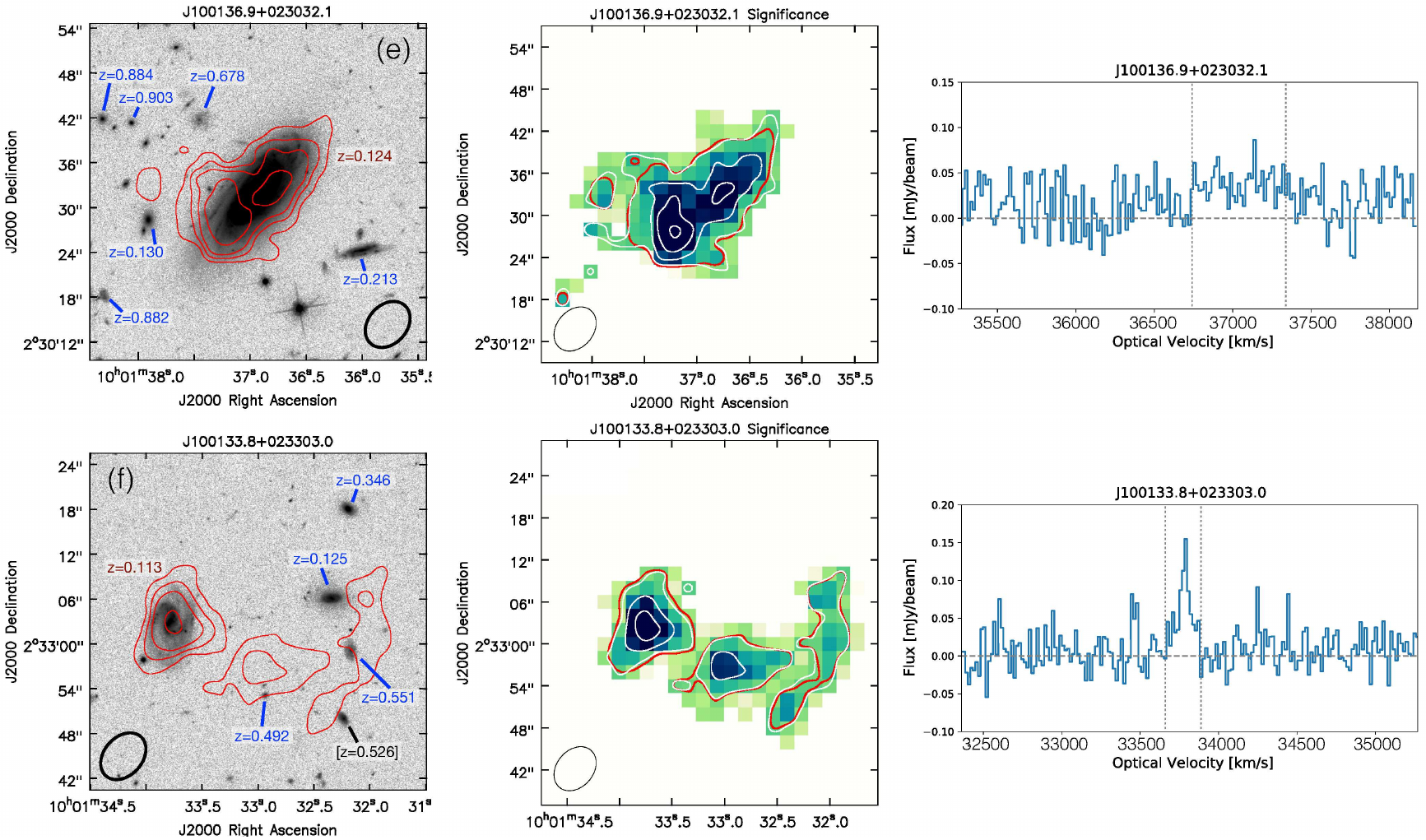}
\caption{Group galaxies in the $z=0.12$ wall. Left: \HI\ contours overlaid on $45^{\prime\prime}\times45^{\prime\prime}$ \textit{HST}/ACS I-band (F814W) images.  Contours are (1.6, 2.7, 5.3, 8, 13, 21)$\times10^{20}$ cm$^{-2}$.  Galaxies with spectroscopic redshifts are in blue (or red if they belong to groups).  Bright galaxies with photometric redshifts are in black and bracketed.  Center: Pixel-by-pixel signal-to-noise ratio. Contours are $\Sigma=(2,4,6,8)$, and the color scale is saturated at $\Sigma=5$. We have included the lowest \hi\ column density contour from the total intensity map in red.  Right: Integrated \hi\ line profile, enclosing lowest contour in left image. The dashed vertical lines indicate the width over which the emission is integrated to calculate the \hi\ mass in Table \ref{properties}. Row (e): a galaxy in Knobel Group 1 (see previous).  Row (f): this galaxy belongs to Knobel Group 296.}
\label{mom0a2}
\end{figure*}
\renewcommand{\thefigure}{\arabic{figure}}

\begin{figure*}
\includegraphics[]{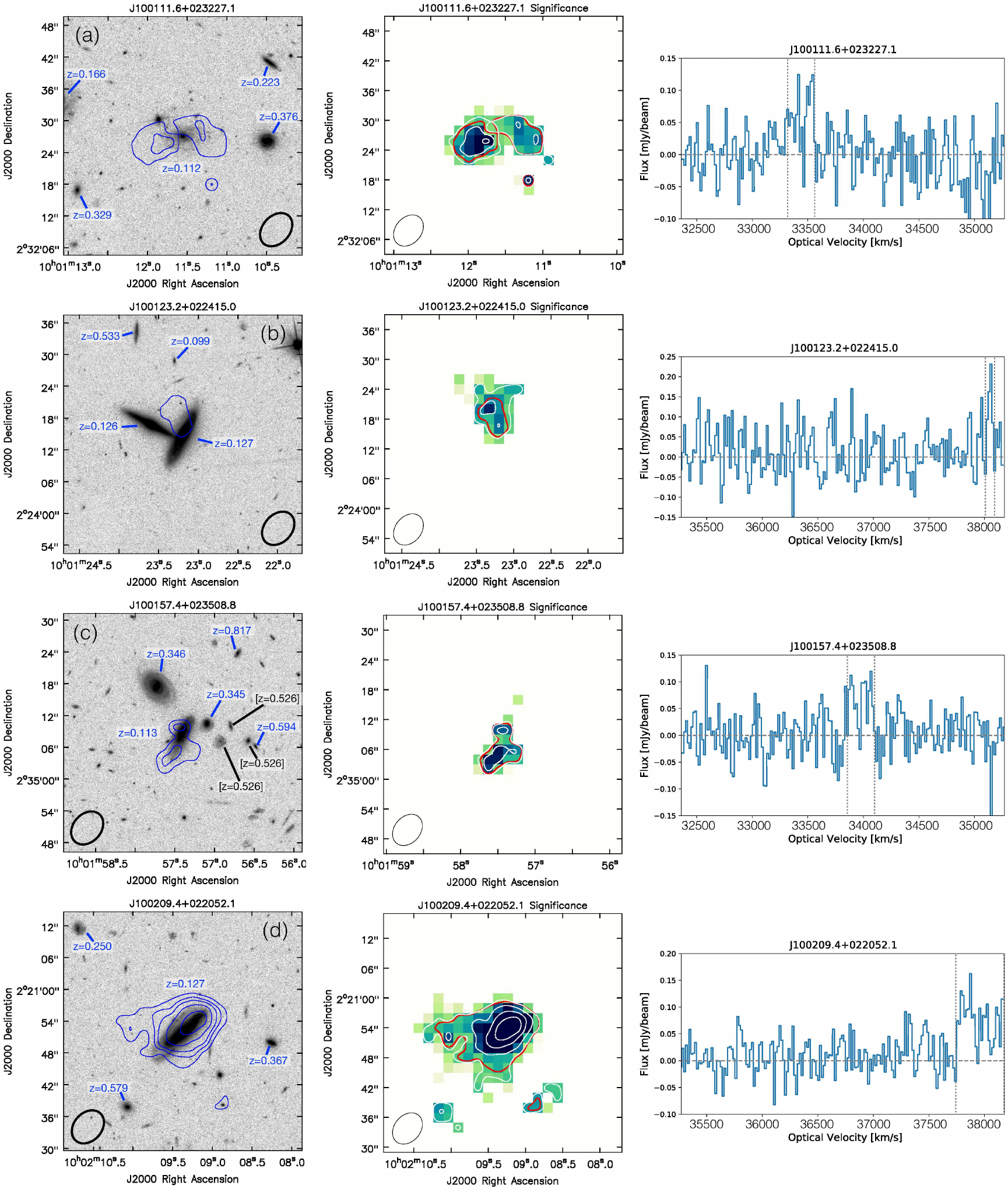}
\caption{Non-group galaxies in the $z=0.12$ wall. Left: \HI\ contours overlaid on $45^{\prime\prime}\times45^{\prime\prime}$ \textit{HST}/ACS I-band (F814W) images.  Contours are (1.6, 2.7, 5.3, 8, 13, 21)$\times10^{20}$ cm$^{-2}$.  Galaxies with spectroscopic redshifts are in blue.  Bright galaxies with photometric redshifts are in black and bracketed.  Center: Pixel-by-pixel signal-to-noise ratio significance maps. Contours are $\Sigma=(2,4,6,8)$, and the color scale is saturated at $\Sigma=5$. We have included the lowest \hi\ column density contour from the total intensity map in red.  Right: Integrated \hi\ line profile, enclosing lowest contour in left image. The dashed vertical lines indicate the width over which the emission is integrated to calculate the \hi\ mass in Table \ref{properties}. For J100209.4+022052.1 whose \hi\ spectrum is presented in panel (d), we searched the adjacent CHILES image cube that spans 1240--1262 MHz and confirmed that there is no galaxy emission below 1260 MHz.}
\label{mom0b}
\end{figure*}

\begin{figure*}
\includegraphics[]{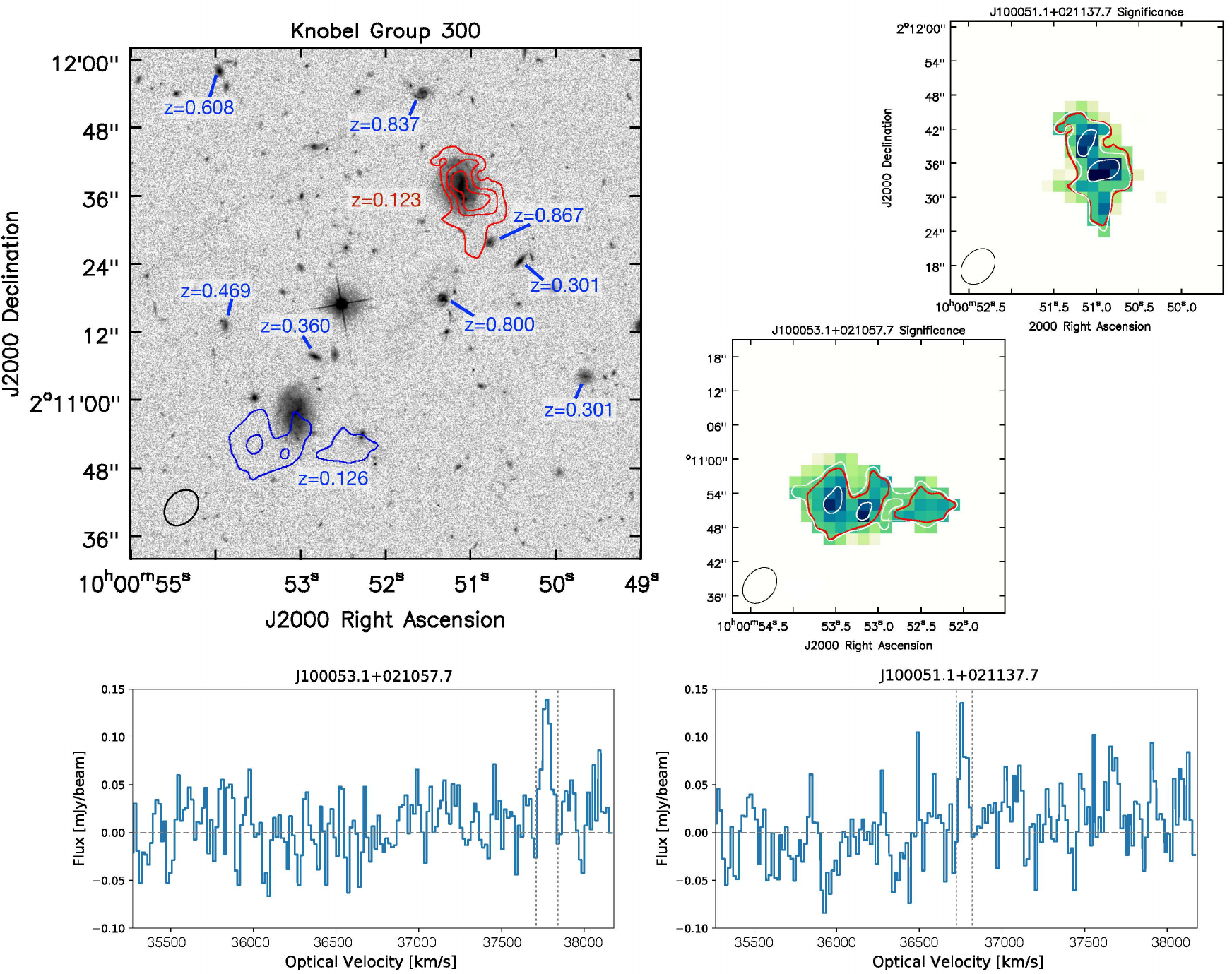}
\caption{Galaxies in Group 300 at $z=0.12$: Top left: \HI\ contours overlaid on $90^{\prime\prime}\times90^{\prime\prime}$ \textit{HST}/ACS I-band image of, from left to right, J100053.1+021057.7 and J100051.1+021137.7. Contours levels are (1.6, 2.7, 5.3, 8, 13, 21)$\times10^{20}$ cm$^{-2}$. Nearby galaxies with spectroscopic redshifts are labeled in blue. Top right: Pixel-by-pixel signal-to-noise ratio significance maps. Contours are $\Sigma=(2,4,6,8)$, and the color scale is saturated at $\Sigma=5$. We have included the lowest \hi\ column density contour from the total intensity map in red.  Bottom: Integrated \hi\ line profile, enclosing lowest contour in top left image.  The dashed vertical lines indicate the width over which the emission is integrated to calculate the \hi\ mass in Table \ref{properties}.  The southern face-on galaxy has an optical redshift of $z=0.1238$, but has an \HI\ redshift of $z=0.1263$, placing it outside the group.  Group membership is discussed in detail in Section \ref{membership}.}
\label{mom0c}
\end{figure*}

\begin{figure*}
\includegraphics[]{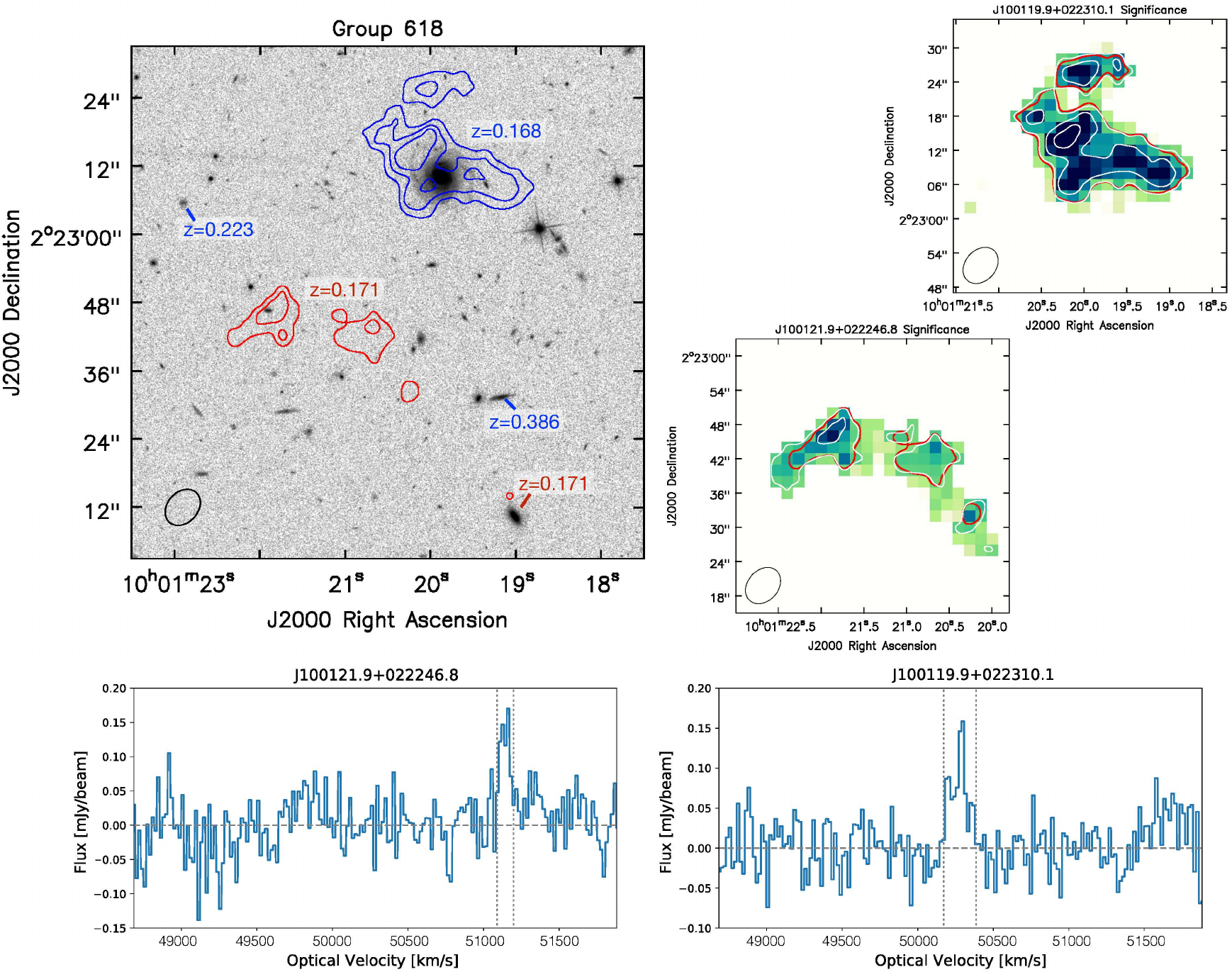}
\caption{Galaxies in Group 618 at $z=0.17$: \HI\ contours overlaid on $90^{\prime\prime}\times90^{\prime\prime}$ \textit{HST}/ACS I-band image of, from left to right, J100121.9+022246.8 and J100119.9+022310.1. Contour levels are (2.0, 3.1, 6.3, 9.5)$\times10^{20}$ cm$^{-2}$.  Galaxies with spectroscopic redshifts are labeled in blue (non-group members) and red (group members).  Top right: Pixel-by-pixel signal-to-noise ratio significance maps. Contours are $\Sigma=(2,4,6,8)$, and the color scale is saturated at $\Sigma=5$. We have included the lowest \hi\ column density contour from the total intensity map in red.  Bottom: Integrated \hi\ line profile, enclosing lowest contour in top left image.  The dashed vertical lines indicate the width over which the emission is integrated to calculate the \hi\ mass in Table \ref{properties}. The large face-on galaxy was a photometric member in \citet{Knobel12} at $z=0.1735$, but has a G10/COSMOS and \HI\ redshift of $z=0.1679$, placing it outside the group. Group membership is discussed in detail in Section \ref{membership}.}
\label{mom0d}
\end{figure*}

\begin{figure*}
\includegraphics[]{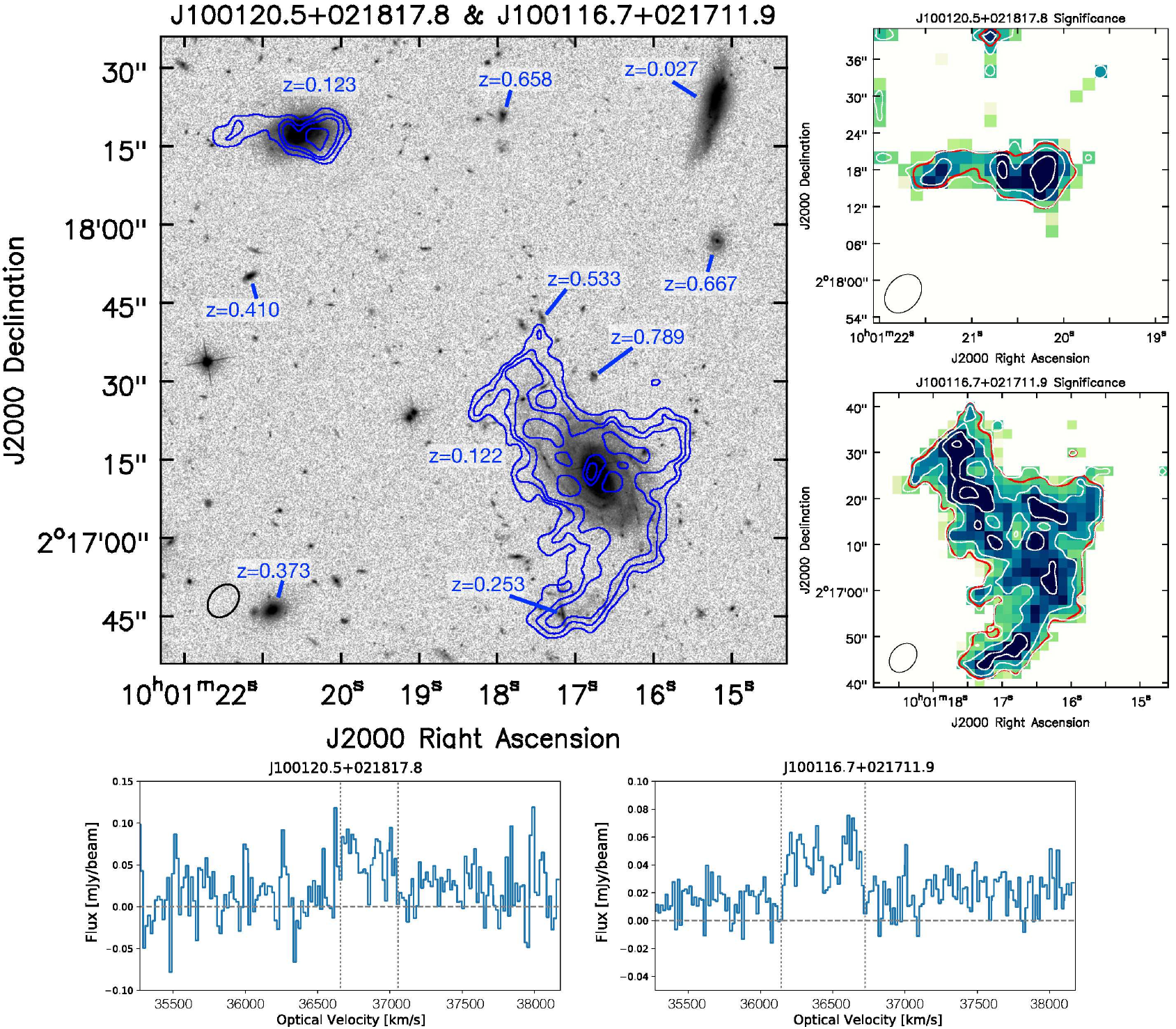}
\caption{Left: \HI\ contours overlaid on $120^{\prime\prime}\times120^{\prime\prime}$ \textit{HST}/ACS I-band image of, from left to right, J100120.5+021817.8 and J100116.7+021711.9.  \HI\ contours are (1.6, 2.7, 5.3, 8, 13, 21)$\times10^{20}$ cm$^{-2}$. Galaxies with spectroscopic redshifts are labeled in blue.  Right: Pixel-by-pixel signal-to-noise ratio significance maps. Contours are $\Sigma=(2,4,6,8)$, and the color scale is saturated at $\Sigma=5$. Bottom: Integrated \hi\ line profile, enclosing lowest contour in top left image.  Bottom: The dashed vertical lines indicate the width over which the emission is integrated to calculate the \hi\ mass in Table \ref{properties}.The galaxies are only separated by $\Delta v\sim420$~\kms.  Their \HI\ morphologies suggest a possible gravitational interaction in the past and are discussed in detail in Section \ref{membership}.}
\label{mom0e}
\end{figure*}

\begin{figure}
\includegraphics[]{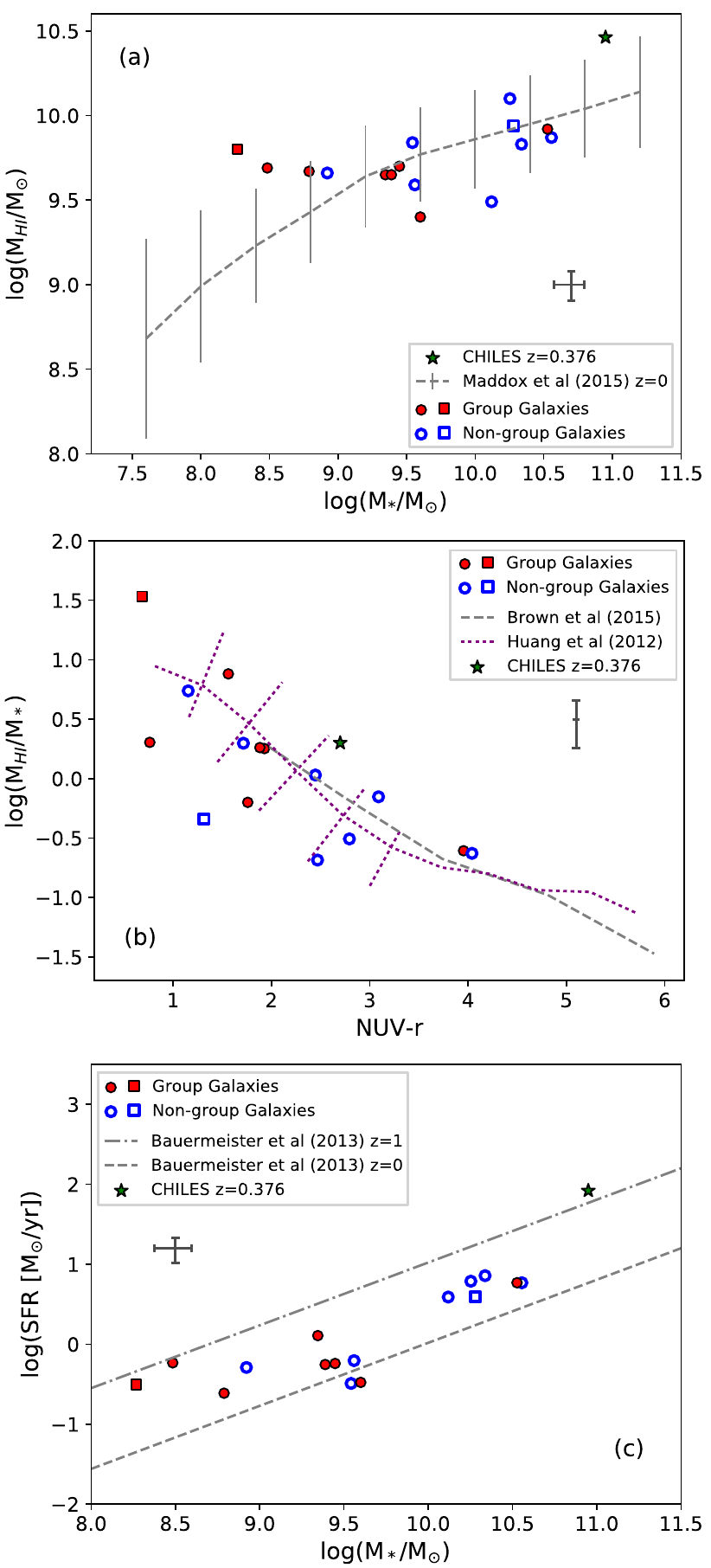}
\caption{(a) \HI\ mass vs stellar mass compared to the median and one sigma spread of the ALFALFA sample at $z=0$ from \citet{Maddox15}.  (b) \HI\ gas fraction vs k-corrected (NUV-r) compared to ALFALFA (\citealt{Huang12,Brown15}). Perpendicular dotted lines represent the spread of the ALFALFA sample where the number of galaxies per bin in \citet{Huang12} is $\ge20$. (c) Star formation rate vs stellar mass compared to the analytic form from \citet{Bauermeister13}. For comparison, we include the highest redshift CHILES detection at $z=0.376$ \citep{Fernandez16}.  
Following Figures \ref{mom0a}--\ref{mom0d}, detections are filled red symbols if they belong to a known galaxy group, and blue empty symbols if they do not.  Square symbols correspond to the sources in the $z=0.17$ volume.  Representative error bars are shown in dark grey, where the uncertainties in the HI mass, stellar mass, and star formation rate are 20\%, 25\%, and 35\% respectively (see text for details). } 
\label{span}
\end{figure}

\subsection{Properties of the \HI-detected galaxies}
\label{props}

Deep multi-wavelength coverage from COSMOS provides powerful ancillary data from which we calculate the overall stellar and star forming galaxy properties. Table \ref{properties} summarizes the measured atomic gas, stellar, and environmental properties of the \HI\ detected galaxies. The columns are as follows: (1) The COSMOS 2008 ID. (2) The G10/COSMOS v05 ID. (3) The optical position from the G10/COSMOS v05 catalogue. (4) The \hi\ redshift from CHILES and the optical redshift from G10/COSMOS v05. (5) The primary beam corrected \HI\ mass which we measure by integrating over the \hi\ line profile presented in Figures \ref{mom0a}--\ref{mom0e}. (6) The width over which we integrate the \hi\ line profile.  This velocity width is in the observer's frame. (7) The stellar mass which we estimate from Spitzer IRAC 3.6 $\mu$m and 4.5 $\mu$m photometric data \citep{Eskew12}, or by using the absolute magnitude derived from a best fit infrared spectral energy distribution, where Spitzer detections are lacking. The best fit infrared SEDs are obtained through the \textit{Le Phare}\footnote{\url{http://www.cfht.hawaii.edu/~arnouts/lephare.html}} \citep{Arnouts99,Ilbert06} with the templates of star forming galaxies \citep{Chary01,Dale01}. (8) The total star formation rates which we estimate by combining UV and IR photometry \citep{Whitaker14}, or when one is missing, from COSMOS 30-band photometry \citep{Ilbert09}.  We find that they are consistent with normal star-forming galaxies. (9) The Knobel group ID. (10) The figure in which we present the total intensity, and significance maps, and \hi\ line profile.

In general we estimate the \hi\ mass to have an uncertainty of order 20\%.  We take a more conservative estimate than the 10-15\% errors typically expected from instrumental uncertainties (e.g.~\citealt{Springob05}) because of the low signal-to-noise nature of our detections.  For barely resolved sources, seen in Figures \ref{mom0a}b, \ref{mom0b}b, and \ref{mom0b}c the uncertainty may be as high as 50\%.  These sources also have a lower peak significance.  As will be discussed, full CHILES will provide a valuable check on the reliability of their \hi\ properties.  The morphology of individual sources are discussed in detail in Section \ref{discussion}.  In addition, we estimate the uncertainty on the stellar mass to be approximately 25\%, and the star formation rate to be 35\%.  The errors on the photometry used to derive these values are very small and underestimate the uncertainty, so we estimate it based on the detailed analysis of SED modeling by \citet{Iyer17}.

Figure \ref{span} illustrates the range of the stellar and \HI\ properties of the detections.  We compare our results with the galaxy population detected in \HI\ by Arecibo Legacy Fast ALFA survey (ALFALFA; \citealt{Haynes11}): the deepest wide-field, blind \HI\ survey at low redshift; and with the highest known \HI\ detection in emission at $z=0.376$: an extremely gas rich, actively star forming spiral detected by CHILES \citep{Fernandez16}. In addition to highlighting group (red) and non-group (blue) galaxies in these plots we also differentiate between the $z=0.12$ detections (circles) and $z=0.17$ detections (squares).

Figure \ref{span}a shows that our \HI\ detections span a broad range of stellar masses, but generally fall within approximately 1 sigma of the median \HI--stellar mass range traced by ALFALFA detected galaxies at a mean redshift of $z\sim0.02$.  The line taken from \citet{Maddox15} describes an approximate ``upper envelope'' in the allowed \hi\ mass for a given stellar mass galaxy.  Aside from the $z=0.376$ source, the most gas-rich outlier is J100121.9+022246.8: the lowest stellar mass galaxy of the $z=0.17$ group detected in \HI\ (middle left of Figure \ref{mom0d}).  The \HI\ appears quite extended for the optical size, and it is possible there may be residual artefacts in the data contributing to an overestimate of the \HI\ mass.

Figure \ref{span}b shows the gas-fraction versus k-corrected NUV-r color of the detections compared to the ALFALFA sample \citep{Huang12} as well as the relation derived from a stellar mass-selected stacked ALFALFA sample \citep{Brown15}.  In general, the majority of our detections are consistent with the measured relations at $z=0$ and the spread seen in the ALFALFA sample.
The galaxies at $z=0.17$ are the greatest outliers on this plot (squares). One object appears to be extremely gas rich for its color: J100121.9+022246.8 (left of Figure \ref{mom0d}) is a very small optical galaxy with an \HI\ tail extending to the southwest. The tail is detected at low significance and, as discussed in Section \ref{membership}, the total \hi\ mass of the galaxy may be overestimated.  However, it is also interesting to note that this galaxy has a high star formation rate for its stellar mass compared (Figure \ref{span}c) to the rest of the \hi\ detections.  This star formation rate may be fueled by a large \hi\ reservoir.
Meanwhile J100119.9+022310.1, the large face-on galaxy at $z=0.17$ (top right of Figure \ref{mom0d}), appears comparatively gas poor for its color.  The northern gas cloud is a continuation of the eastern spiral arm, but deeper observations are likely to reveal even more emission to the south that was too faint to be well characterized by the SoFiA source finder.  

Finally, Figure \ref{span}c shows the star formation rate versus stellar mass compared to the analytic relation for star-forming galaxies from \citet{Bauermeister13} at different redshifts.  Two group member galaxies, J100121.9 noted above, and J100126.9 (Figure \ref{mom0a}c), have the highest star formation rates for their stellar mass and are actually more consistent with the $z=1$ empirical relation of \citealt{Bauermeister13}.  These two galaxies also fall above the $M_{HI}$ versus $M_{\odot}$ relation of \citet{Maddox15}, suggesting that even if we suspect that we are overestimating their mass because of the contribution of low significance \hi\ emission, large \hi\ reservoirs may be fueling their relatively more enhanced star formation.  This is also what makes the CHILES detection at $z=0.376$ unusual: its specific star formation rate is closer to galaxies at $z>1$ \citep{Fernandez16}.  The rest of the CHILES detections between $0.11<z<0.18$ are also on average a bit higher than the SFR versus stellar mass relation for their redshift, but given the spread in the relation, the small number statistics, and the fact that the galaxies are massive in \hi\, we cannot conclude that they are an unusual population.

We also note from Figure \ref{span}c that non-group galaxies tend on average to have higher stellar masses than group galaxies.  This may be because the group population is dominated by a low mass satellite galaxy population which live around a comparatively more massive central galaxy (e.g.~\citealt{Feldman10}), while the non-group galaxies are by definition the most massive galaxy in their dark matter halo, but we lack the sample size to say for certain that this is the cause of the stellar mass segregation in this plot. 

In short, despite the fact that we are only sensitive to the most \HI\ massive part of the galaxy population, we find that, at $z=0.12$, the \HI\ detections appear to follow the expected trends in their star formation versus stellar mass for their redshift and, as a population, show no obvious difference in their overall gas content from the $z=0$ ALFALFA sample.  This is probably not surprising: at $z=0$ \HI\ mass is only weakly correlated with stellar mass above $10^9$~\msun\ \citep{Maddox15}, and we do not expect much evolution in the \HI\ content between $0<z<0.2$.  At $z=0.17$, the J100121.9 appears to be slightly gas rich with respect to the rest of the sample and in relation to its stellar content and color.  As is suggested in Figure \ref{3d}, and will be discussed in the following sections, this galaxy lives on average in a lower density environment than those galaxies at $z=0.12$.

\section{Discussion}
\label{discussion}

The majority of the \HI\ detections in our sample show extended and/or asymmetric morphologies.  To understand the \HI\ morphology, we examine the environment in which the galaxies live.  We consider two different methods of quantifying environment: the groups to which galaxies belong based on a FoF group finding algorithm; and their location within the filamentary structure based on the DisPerSE topological algorithm.  In addition to these measures, which primarily rely on spectroscopic redshifts, we visually inspected $90^{\prime\prime}\times90^{\prime\prime}$ \textit{HST} images around each \HI\ detection to look for signs of interactions with companions including objects with only photometric redshifts. This corresponds to roughly $195\times195$ kpc at $z=0.12$ and $262\times262$ kpc at $z=0.17$.

\subsection{\HI\ Morphology in the Context of the Group Environment}
\label{membership}

Within the $z=0.12$ wall, 7 of 14 \HI\ detections belong to three Knobel groups (summarized in Table 2).  These galaxies exhibit a wide range of \HI\ morphologies.  At $z=0.17$, both \hi\ detections were associated with either a spectroscopic or photometric member of one Knobel group.  Nearly all \HI\ detected group members (6/8) show extended emission at 1.3-3 times the extent of the stellar disk in the \textit{HST} I-band images, after correcting for beam smearing.  (Our estimate assumes an underlying flat distribution of \hi, convolved with the Gaussian beam.)  Given the variety and small number statistics, it is difficult to make strong general conclusions.  Below, we discuss \HI\ detections within individual groups.

\textbf{Group 1} --- This is the most massive group in the entire CHILES volume.  It has 30 confirmed optical spectroscopic members and 26 candidate members based on photometric redshifts in the Knobel catalogue.  We detect five member galaxies above an \HI\ mass of $10^{9.6}$~\msun\ (Figure \ref{mom0a}).  Four of the \HI\ detections coincide with known spectroscopic members, and one \HI\ detection provides a confirming spectroscopic redshift for a photometric candidate (J100131.5).  Three of the five \HI\ detections show extended emission based on their higher significance emission.  J100119.8 and JJ100126.9 are resolved by $\sim$2.5-3 beams, and J100136.9 is well resolved covering an area of $\sim$12 beams.  The two galaxies without extended emission are J100120.3 (Figure \ref{mom0a}b) and J100131.5 (Figure \ref{mom0a}d).  The former is only very marginally resolved, and has an \HI\ distribution that does not extend beyond the stellar disk.  One possibility is that this galaxy has a truncated \HI\ disk, because it has been living in Group 1 for a longer period of time and undergone multiple gravitational interactions or has had the gas in the outer disk removed by the intragroup medium.  The latter has a very disturbed optical morphology and may be currently undergoing a merger.  It is well resolved in \hi.

Indeed, we note that J100131.5 (Figure \ref{mom0a}d) is actually detected as two galaxies in the COSMOS 20k and G10/COSMOS catalogues.  It could be two galaxies observed in projection, however, given that the optical morphology of the main galaxy at $z_{\texttt{spec}} = 0.121$ looks disturbed; that the photometric redshift of the other galaxy candidate is very close ($z_{\texttt{phot}}=0.13$); that the \HI\ appears to trace a common envelope around them; that the peak of the HI intensity map falls on the object which only has a photometric redshift; and that the integrated profile of the \hi\ envelope covering both sources is very narrow, we believe this is an excellent candidate for a late stage merger within the group.  As a result, in Table \ref{properties}, we treat it as one object and combine the derived stellar mass and star formation rates for the two sources assuming the same spectroscopic redshift.

\textbf{Group 296} --- We detect the brightest group galaxy (BGG) of Group 296 which has the most disturbed \HI\ morphology of all our detections.  An \hi\ tail is well resolved and appears to be very extended to the west  (Figure \ref{mom0a}f).  The two other spectroscopic confirmed group members lie to the east outside the image and are 1 and 2 orders of magnitude lower in stellar mass.  Only one object coincident with the tail has a spectroscopic redshift below $z=0.3$, and it is still separated by $\Delta z=0.022$ ($>6000$~\kms), so the origin of the \HI\ tail is unclear.  It may be the result of an interaction with another unidentified member, or it may be due to noisy data as this source was characterized by the less stringent pixel dilation method in SoFiA.  

\textbf{Group 300} --- We detect two out of three galaxies in Group 300, including the BGG (upper right of Figure \ref{mom0c}).  However we measure an \hi\ redshift for J100053.1 that suggests it is not in fact a group member.  In the original Knobel 20k catalog, this galaxy had a PA of 0.85.  The \hi\ redshift puts it roughly 660~\kms\ further away from the BGG.  The two roughly face-on galaxies in Figure \ref{mom0c} both have \HI\ extending to the south, but since the galaxies are separated by $\sim$2000~\kms, it is unlikely they have interacted with one another.  The last member of the group is more than an order of magnitude smaller in stellar mass, and lies to the north outside of the image.  While J100053.1 appears to be quite asymmetric, continuous low level emission to the north is visible by-eye in the data cube but is not picked up by the source finder.  Both sources are resolved with $\sim$3-4 beams along their \hi\ major axis.

\textbf{Group 618} ---  At $z=0.17$, we detect one out of two spectroscopic members and one photometric candidate.  All three galaxies are visible in Figure \ref{mom0d}.  The photometric candidate is a large face-on spiral (J100119.9), but its \HI\ redshift strongly suggests it does not belong to the group as it differs by $\Delta z=0.003$ ($\sim$900~\kms) from the other two members.  The galaxy has an \hi\ cloud to the north that is an extension of the eastern spiral arm, and deeper observations will likely reveal more \hi\ emission to the south that was too faint to be picked up by the source finder.  J100121.9 is a very small optical galaxy, which shows an extension to the southwest towards the \hi-undetected group member.  Even after taking into account beam-smearing, it appears that the \hi\ presumed to be in the disk of the galaxy is quite extended with respect to the optical disk. As mentioned in Section \ref{props}, it is quite gas-rich relative to ALFALFA and to the rest of this CHILES sample.  We may be overestimating the \hi\ mass, or it may in fact have a large gas reservoir fueling its star formation.



\begin{figure*}
\includegraphics[width=\textwidth]{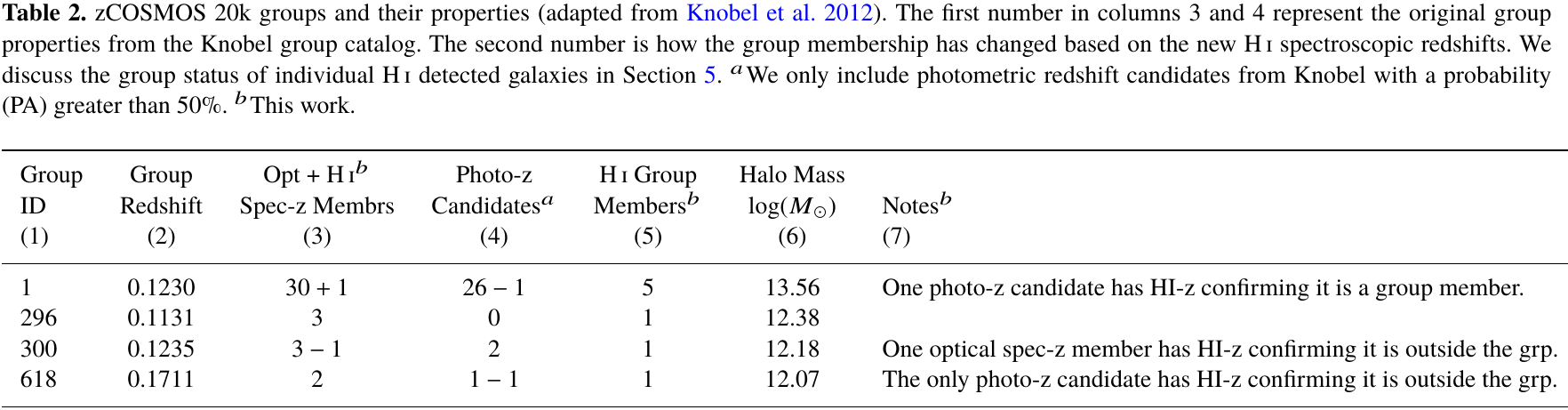}
\label{grouptab}  
\end{figure*}

\subsection{Non-group Galaxies}
\label{ngg}

The reprocessing of zCOSMOS survey data has increased the number of available redshifts and their reliability since the 20k group catalogue was created.  As discussed in the previous subsection, this has led to J100131.5 being spectroscopically confirmed as a member of Group 1, and J100053.1 and J100119.9 most likely being eliminated as a members of Groups 300 and 618.  A full re-evaluation of the Knobel group catalogue is outside the scope of this paper, but a new group catalogue including improved redshift measurements is an ultimate goal of the G10/COSMOS project (e.g.~\citealt{Davies18}).

In examining Knobel non-group members we inspected the stellar morphology of \HI\ detections and nearby galaxies in COSMOS \textit{HST} mosaics, and their spectroscopic and photometric redshifts, to search for new companions.  In doing so, we kept in mind that for pairs of galaxies in \citet{Knobel12}, the FoF linking lengths were 0.225 Mpc and approximately 260~\kms.  Among the non-group galaxies (blue \HI\ contours in Figure \ref{mom0b}--\ref{mom0e}), five of eight are extremely asymmetric in their \hi\ distribution compared to relatively well defined stellar disks.  We discuss in detail below their morphology and environment.

\textbf{Disturbed non-group galaxies with candidate companions ---}  Three \HI\ detections have nearby companions with spectroscopic redshifts that are far enough away to not be considered groups by the FoF algorithm.  

J100123.2 (Figure \ref{mom0b}b) has an edge-on companion very close in projection.  In this case, the spatial separation is 6 arcseconds, but the velocity separation is 353~\kms.  While outside the definition of a group or pair in the Knobel catalogue, the \HI\ morphology suggests that they have undergone an interaction in which the gas is being pulled from one of the galaxies, while the other was undetected in \HI\ at our current limits.  The \hi\ we detect in this galaxy is only very marginally unresolved, and we may only be seeing the highest density gas in the system.

Figure \ref{mom0e} shows two more non-group galaxies with disturbed morphologies, J100120.5 and J100116.7.  Both are massive with $10^{9.9}$ and $10^{10.4}$~\msun\ of \HI, respectively, and they are both well resolved in the \hi\ observations.  The galaxies are separated by approximately 1.3~arcmin ($\sim$170 kpc) and 420~\kms\ (Figure \ref{mom0e}).  Again, the velocity difference is too great to be considered a pair in the group catalogue.  However, based on their positions and relative velocities, they may have undergone an interaction in the last 40 Myr, which may explain their \HI\ morphologies if they underwent a tidal interaction leading to tidal tails on opposite sides of both galaxies (e.g.~\citealt{Toomre72,Hibbard96}).  The tidal tail south of J100116.7 has a faint stellar component which is continuous along the length of the tail.  The edge-on galaxy that is superimposed on the end of the tidal tail is a background object with a spectroscopic redshift that places it at twice the distance of J100116.7. Both of these \hi\ detections in Figure \ref{mom0e} are well resolved, and their morphology is relatively more significant.

There are an increasing number of cases where high speed encounters (relative to the typical group velocity dispersion) between galaxies are suspected to have produced a dramatic effect on the gas distribution while the stellar component appears relatively undisturbed (e.g.~NGC~4254 in the Virgo Cluster, \citealt{Duc08}; NGC~3162 with HCG~44, \citealt{Leisman16,Hess17}).  The relative velocity and spatial separation between these systems suggest they are not gravitationally bound in their own subhalo: their past interaction is only revealed through the presence of \HI\ tails, sometimes extending beyond 300 kpc.  It will be interesting to see if deeper images with the full CHILES sensitivity reveal even more extended \HI\ tails.

\textbf{Disturbed non-group galaxies without companions ---} The fourth galaxy to show extreme asymmetry is J100157.4 (Figure \ref{mom0b}c).  Interestingly, the stellar distribution shows asymmetry in the same direction as the \HI\ with a lower surface brightness component extending to the southeast, relative to the bright central disk.  We caution however that the \hi\ is only marginally resolved and the clean beam is elongated in a similar orientation to the galaxy.  There are several galaxies seen close in projection, but they appear to belong to background collections at $z\sim0.346$ or $z\sim0.526$.  Unfortunately, we have not identified a clear candidate for a recent tidal interaction.

\textbf{Other non-group galaxies ---}  Of the remaining two non-group members, we note that J100111.6 in particular has quite an extended \HI\ envelope compared to the optical extent that is visible in the \textit{HST} image (Figure \ref{mom0b}a).  As we will show in Section \ref{hifilament}, it is also the most isolated of our \HI\ detections.  Finally, J100209.4 (Figure \ref{mom0b}d) is massive in both its \hi\ and stellar components, but not unusually so relative to other \hi\ detected samples.  It is a strong \hi\ detection with a hint of low level emission outside the stellar disk.  Both these non-group galaxies are resolved with 3 beams across the major axis.

\begin{figure*}
\includegraphics[]{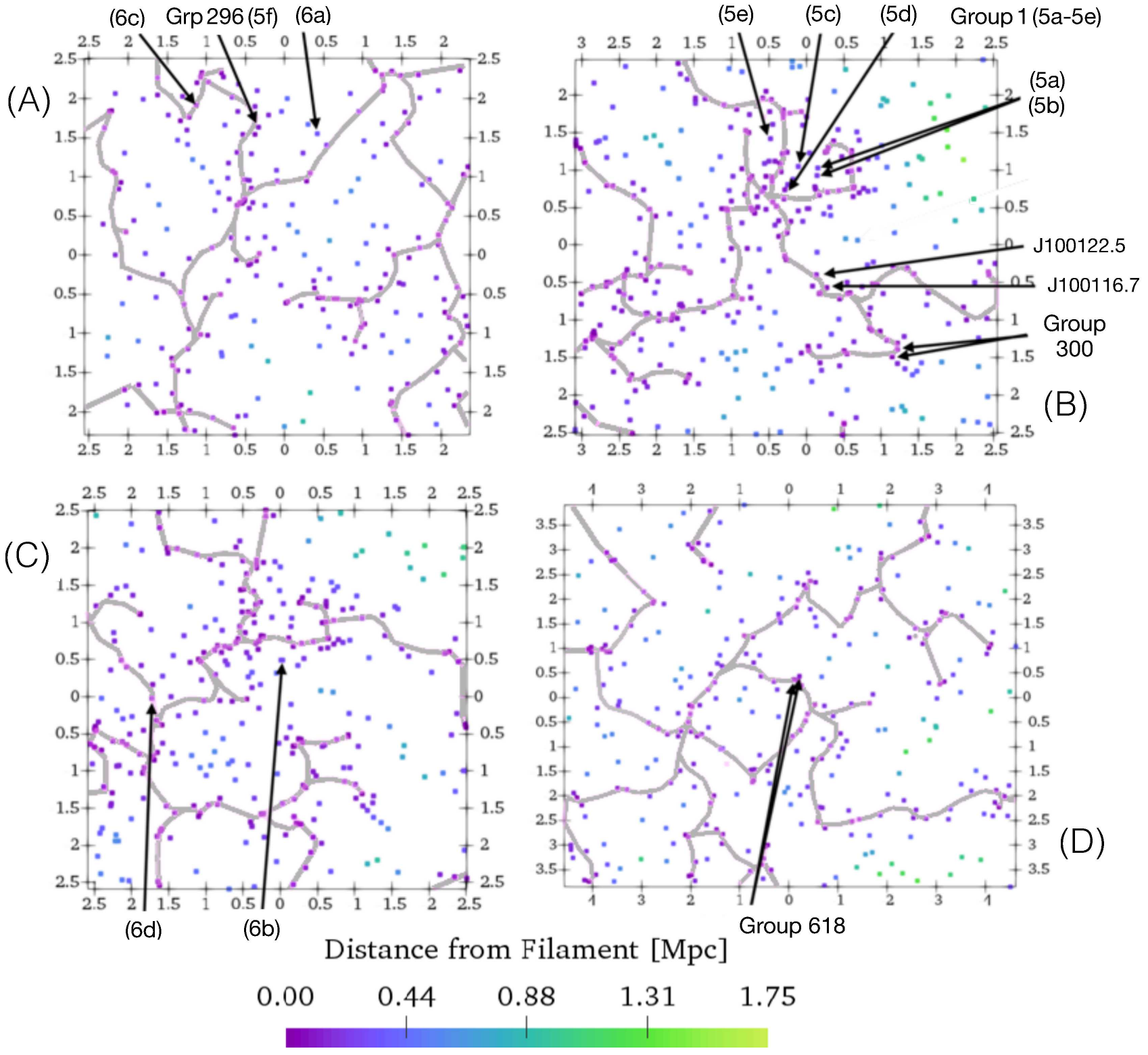}
\caption{The filamentary network of the cosmic web overlaid on the distribution of galaxies in the following redshift ranges: (A) $z = 0.107 - 0.117$, (B) $z = 0.118 - 0.128$, (C) $z = 0.121 - 0.131$ and (d) $z = 0.165 - 0.175$. The group \HI\ detections are indicated based on their row number in Figure \ref{mom0a}(a-f), their Knobel group number (as in Figures \ref{mom0c} and \ref{mom0d}), or for non-group \HI\ detections, by their row number in Figure \ref{mom0b}(a-d), or their right ascension (as in Figure \ref{mom0e}. The redshift ranges were chosen such that the \HI\ detections within each panel have a redshift that is more than $\Delta z=0.0025$ from the boundaries for that panel. The color scale corresponds to the distance of each galaxy to the nearest filament. The spatial axes are in Mpc.}
\label{disperse}
\end{figure*}

\subsection{Filaments and Voids}
\label{filamentvoid}

A different, but complementary, way to look at environment is to consider a galaxy's location within the cosmic web.
Figure \ref{disperse} shows the filamentary network measured by DisPerSE in four redshift slices which contain the \HI\ detections.  As described in Section \ref{dispersedesc}, the width of the slices used to calculate the positions of critical points from the galaxy density field was $\Delta z=0.01$.  However, we only show galaxies that are within the central $\Delta z=0.005$.  This allows us to be more inclusive of the 3D structure despite collapsing it into two dimensions, but we only calculate the distances to structures for galaxies that are well within it.  A more complete description of the trade off between the 2D and 3D implementation of DisPerSE will be discussed in Luber et al (in preparation).  A description of the two methods can be found in \citet{Sousbie11,Sousbie11b}.  We also calculate the structure over the full $2\times2$ deg$^2$ COSMOS field but only show the $\sim$0.5 $\times$ 0.5 deg$^2$ CHILES field of view.  In this way, the calculation of structure in these insets are not impacted by edge effects.

Based on the extracted filamentary structure, we see that the wall noted in redshift slices at $z=0.12$ (Figure \ref{3d}, or Figure 2 in \citealt{Fernandez13}) is dominated by the high density of galaxies in the upper middle region of Panels (B) and (C) of Figure \ref{disperse}.  This high density of galaxies is at the intersection of two filaments which run roughly north-south and east-west in Panel (B).  We see that the general orientation of these filaments persists in Panel (C), giving a sense of its extension in redshift space, although the filament extending to the south has disappeared.  In fact, the intersection of these three or four filaments is the location of Group 1, identified by the Knobel FoF group-finding algorithm.  This narrow redshift wall at $z=0.12$ also contains several more small ($N=2$ -- 4) Knobel groups (Figure \ref{3d}).

By comparison, Panels (A) and (D) in Figure \ref{disperse} suggest that regions of low galaxy density exist at $z=0.11$, immediately in front of the $z=0.12$ wall but still within the redshift range of the 1260--1282 MHz data cube, and in the $z=0.17$ volume, respectively.  The overall number density of galaxies in these two slices is lower, and the extracted filaments do not appear to be as densely woven as in Panels (B) and (C).  This is consistent with the 3D representation of the data in Figure \ref{3d}.  In particular, the small galaxy groups at $z=0.17$ contain only two or three members and do not cluster strongly in redshift space as they do at $z=0.12$.

\subsection{\HI\ in Filaments}
\label{hifilament}

Figure \ref{disperse} also indicates the location of the \HI\ detections relative to the filaments.  The first clear result is that, at our \HI\ mass sensitivity, none of the \HI\ detections are among the most isolated galaxies found by the algorithm.  In fact, the majority (9/16) of the \HI\ detections appear to lie on the backbone of filamentary structure, including three group member galaxies, and six non-group members. These galaxies are all within 100 kpc of a filament.  The three group members all live in small groups made up of two to three spectroscopic members (Groups 296, 300, 618).

Even more intriguing is the \HI\ morphology of the non-group members which lie on filaments.  Three of the six galaxies show interesting alignment between their \hi\ morphology and orientation of the filaments in which they are embedded.  For example, J100157.4 (Figure \ref{mom0b}c) has no known companions, but has an asymmetric \hi\ tail and stellar disk which are extended parallel to the filament in Panel (A) of Figure \ref{disperse}, although we noted in Section \ref{ngg} that the emission is detected at low significance and marginally resolved.  
Meanwhile, J100120.5 and J100116.7 (Figure \ref{mom0e}) lie on a filament which is aligned parallel to the line connecting them and the axis along which they may have previously interacted, as evidenced by their opposing tidal tails discussed in Section \ref{ngg}.  The alignment is seen in the high significance contours.  We also acknowledge that these two cases of alignment with a filament may be coincidence and more studies of resolved \HI\ galaxies with respect to the large scale environment are required.

Those galaxies that do not lie directly on a filament potentially live in greater isolation.  However, of the seven galaxies which do not lie on a filament, five of them belong to Group 1 (Figure \ref{disperse}B).  The output of DisPerSE is best at defining filaments, walls, and voids, rather than clusters, but by combining it with the group catalogue we identify structure similar to what is seen in cosmological simulations, where the most massive group in our volume lies at the intersection of two filaments.  Thus, these five galaxies live in a high density environment.  Of the remaining two galaxies J100123.2 ((b) in Figures \ref{mom0b} and \ref{disperse}) was shown to have likely undergone an interaction with an edge-on companion 353~\kms\ away from its current position, so is not particularly isolated. 

Finally, J100111.6 ((a) in Figures \ref{mom0b} and \ref{disperse}) remains as the most isolated galaxy in the sample.  Interestingly, it also shows the most extended, and relatively symmetric \hi\ distribution of the sample.  It has a low star formation rate relative to other \HI\ detections at $z=0.11$--0.12, and a high relative \HI-to-stellar mass fraction, falling just above the 1$\sigma$ range from ALFALFA (Figure \ref{span}a).  Taken together, this galaxy appears to be less disturbed and producing stars more passively than galaxies in the wall.  Its evolution may be regulated by secular rather than environmental processes.  This is consistent with the findings of void galaxy studies which find galaxies in low density regions have more extended \HI\ envelopes \citep{Kreckel11,Kreckel12}.

Overall, we find that there is not an obvious one-to-one matching of filamentary structure to group/non-group environment.  Filaments contain both group and non-group galaxies.  However, we do find that the majority of \hi\ detections, at our relatively high \hi\ mass limit, live within $\sim$250 kpc of a filament, and several galaxies which do not belong to a group show an interesting correlation with their \hi\ morphologies and the alignment of the filaments in which they live.

\subsection{Future Work}

In this and previous work at $z=0.376$ \citep{Fernandez16}, we have shown that we resolve extended structure in \hi-detected galaxies.  Although the total intensity maps have been created by SoFiA in order to be objective, unbiased, and easily reproducible, they may still be impacted by the noise at low column densities, and we acknowledge that our interpretation of asymmetry has been somewhat subjective.  In order to assess the reliability of our total \hi\ intensity maps, we also provide significance maps, and we propose that these should become a standard output for future \HI\ surveys.  Deeper images which will come from combining additional CHILES data will provide even more reliable morphological information.
In the future, when the impact of image resolution and how to correct for the noise is better characterized, we plan to objectively quantify morphology of the CHILES \hi\ detections as in \citet{Giese16}.
Ultimately, the comparison between the intensity maps presented here with those eventually derived from the full survey data will inform future deep \hi\ surveys with SKA precursors as to the reliability of interpreting morphology from low column density gas in resolved \HI\ objects on the cusp of detection.

\section{Summary}
\label{summary}

We have presented an analysis of the overall data quality in a volume of the CHILES survey most strongly impacted by satellite and ground based radio frequency interference.  CHILES Epoch 1 consists of the first 178 hours of observing, and the frequency range considered here accounts for less than 10\% of the total bandwidth.  
We identify by-eye \hi\ candidates with known spectroscopic redshifts and conducted detailed testing of the automated spectral source finder SoFiA, which we used to confirm \HI\ detections with a high degree of confidence.  Unfortunately, not all sources are detected with a single run of the source finder, suggesting that further optimization is required before it can be run blindly.  This is true even when the search volume is limited to objects with known optical spectroscopic redshifts.  Fully blind runs of the source finder miss the faintest sources if the signal-to-noise detection threshold is set too high; however lowering the threshold too much results in many false detections.

We detect a total of 16 \hi\ sources with 14 sources residing in the data cube spanning 1260--1282 MHz ($0.108\le z\le 0.127$), and 2 sources residing in the cube spanning 1200--1222 MHz ($0.162\le z\le 0.183$).  The \hi\ mass sensitivity of the images from Epoch 1 of the CHILES survey is fairly high, measuring just below the knee of the HIMF at $z=0.12$, and at or above the knee at $z=0.17$.  We find that the galaxy population at these redshifts does not differ significantly from $z=0$, although we do suffer from low number statistics.

We discuss in detail the \HI\ morphologies of our detections in the context of their local group and global filament/wall/void environments. In addition to \hi\ total intensity maps, we presented significance maps which are critical in assessing the reliability of the \hi\ morphology of low signal-to-noise detections. We propose that these become common place for resolved \hi\ surveys in the SKA pathfinder era and beyond.

We also demonstrated that multiple environmental measures, such as a group catalogue combined with a measure of the larger-scale filamentary and void structure, provide a more complete picture in which to interpret the observed \HI\ morphologies.  In general, groups tend to have a higher galaxy density than filaments, and the group-finding algorithm uses three dimensional information to characterize the galaxy environment.  By comparison, the filaments give a sense of the larger scale environment, but in order to interpret the position of galaxies within them, the relationship between galaxies and the environment is quantified in one or two dimensions, such as distance to the nearest filament, or by qualitatively evaluating the web of filaments in a narrow redshift slice.  

In the context of the group environment, we observe a broad range of \hi\ morphologies.  Extended, asymmetric morphologies may be evidence of past interactions within the group.  However, even non-group galaxies show extended or asymmetric \hi\ distributions.  We searched for nearby companions of non-group galaxies and found that half of them have likely had a recent interaction with a nearby companion that was not close enough to be linked by the FoF group-finding algorithm.  Thus, while group catalogues may have clustering or halo occupation properties which match with expected group populations, on an individual basis they are a poor characterization of groups with small ($N=2$ -- 4) membership.  Galaxy samples like this subset of CHILES greatly benefit from a closer look to understand the history of interactions revealed by \hi\ observations.  None of the interactions, between group or non-group members, are obvious from the optical imaging alone.  By comparing the spectroscopic and photometric members from the Knobel group catalogue with our \hi\ redshifts, we were also able to confirm one photometric member of Group 1, found that one spectroscopic member of Group 300 may be a background galaxy, and that a photometric member of Group 618 is more likely a foreground galaxy.

We used DisPerSE to look at galaxy environment in the context of large-scale filaments which characterize the cosmic web.  We found that all but two \hi\ detections either lie within 100 kpc of a filament, or are members of the most massive galaxy group at the intersection of two filaments.  Several non-group galaxies exhibit an intriguing alignment between their \hi\ morphology and the orientation of the filament in which they are embedded, although small number statistics prevent us from making any strong conclusions about how the filaments may impact the \hi\ distribution in the galaxies.

Finally, DisPerSE combined with the group catalogue allowed us to identify the most isolated galaxy in our sample.  We expect to find more such cases as we work towards completion of the full CHILES survey.  In general, however, we find that none of these \hi\ massive detections are very isolated.

Epoch 1 image cubes contain only 18\% of the total observing time of the CHILES survey, so the interpretation presented here may be subject to revision.  The data quality is good and we demonstrate that it is still possible to detect \hi\ sources in regions of the spectrum impacted with RFI by employing careful flagging.  
With deeper images from the full CHILES survey, we will be able to review the \HI\ intensity maps and provide insight for future resolved deep \HI\ surveys with MeerKAT and the SKA which will seek to go to lower \HI\ masses and column densities at increasingly high redshift.

\section*{Acknowledgements}

The research of K.M.H.~and J.M.v.d.H.~are supported by the European Research Council under the European Union's Seventh Framework Programme (FP/2007-2013)/ERC Grant Agreement nr.~291531.  CHILES is supported by NSF grants AST-1413102, AST-1412578, AST-1412843, AST-1413099 and AST-1412503. X.F.~is supported by an NSF-AAPF under award AST-1501342. Support for M.S.B.~is provided by the NSF through the Grote Reber Fellowship Program administered by Associated Universities, Inc./National Radio Astronomy Observatory. K.K.~gratefully acknowledges support from grant KR 4598/1-2 from the DFG Priority Program 1573.  We thank Tom Oosterloo for comments on an earlier version of this paper which have led to improvements.  We thank the anonymous referee and the scientific editor for their comments which improved the quality of this paper.

The interactive 3D figure was generated using the Mayavi package in Python \citep{Ramachandran12} and was converted to u3d format for inclusion in latex file using DAZ Studio.  We greatly benefitted from the guidance of the ``3D interactive graphics in PDF Tutorial'' by J.~Peek on AstroBetter.  This research made use of Astropy, a community-developed core Python package for Astronomy \citep{Astropy13}.  Scatter plots were made using Matplotlib \citep{Hunter07}.  

Parts of this research were conducted by the Australian Research Council Centre of Excellence for All-sky Astrophysics (CAASTRO), through project number CE110001020.

The National Radio Astronomy Observatory is a facility of the National Science Foundation operated under cooperative agreement by Associated Universities, Inc.

The Amazon Web Services were provided with credits from the AstroCompute program, managed by the SKA Office.

The G10/COSMOS redshift catalogue 
uses data acquired as part of the Cosmic Evolution Survey (COSMOS) project and spectra from observations made with ESO Telescopes at the La Silla or Paranal Observatories under programme ID 175.A-0839. 
Full details of the data, observation and catalogues can be found in Davies et al.~(2015) 
or on the G10/COSMOS website: cutout.icrar.org/G10/dataRelease.php.

This research has made use of the NASA/IPAC Infrared Science Archive, which is operated by the Jet Propulsion Laboratory, California Institute of Technology, under contract with the National Aeronautics and Space Administration.








\appendix

\section{Position-Velocity Maps}

\begin{figure*}
\includegraphics[]{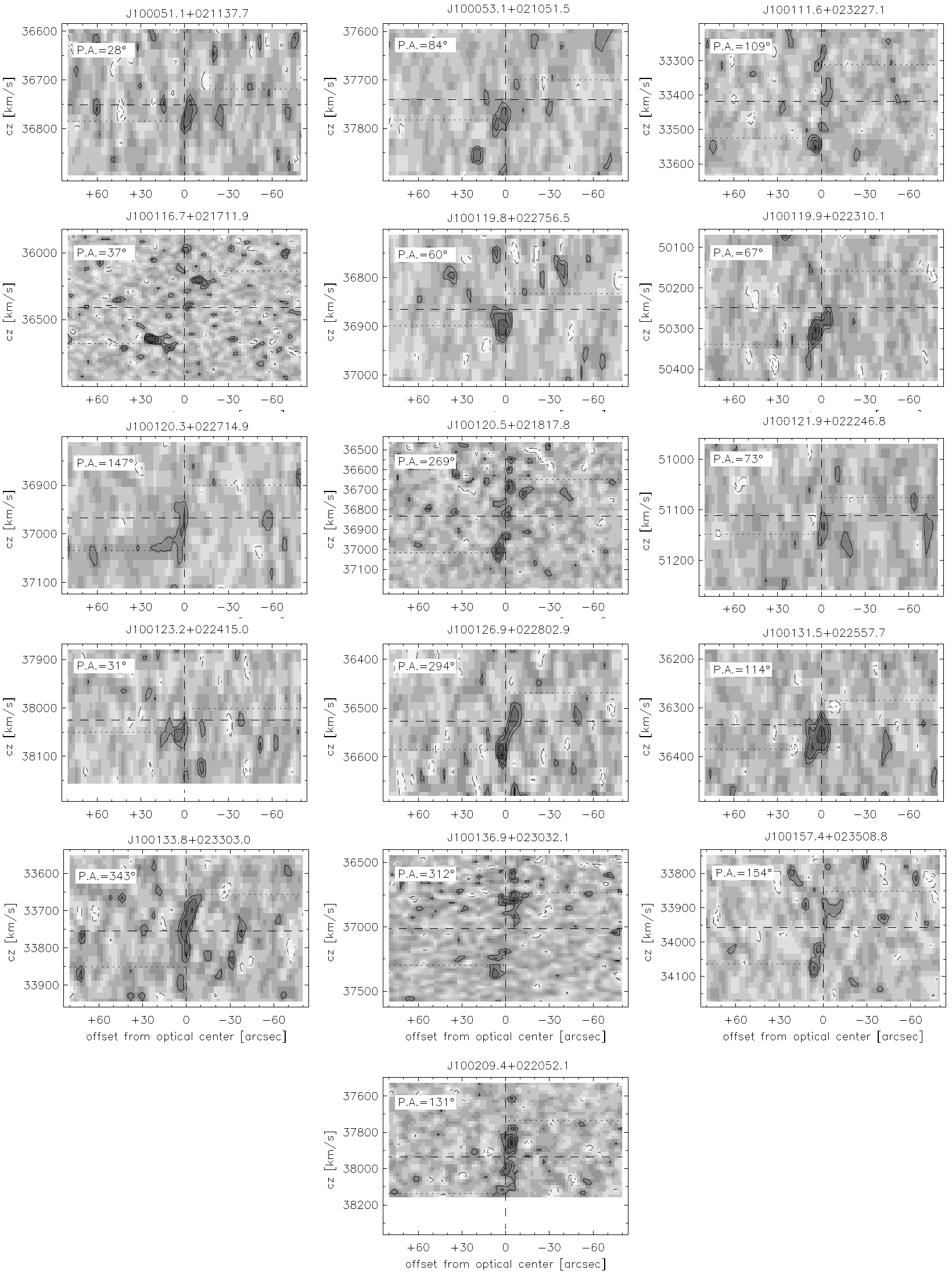}
\caption{Position-velocity maps of the \hi\ detected galaxies.  The galaxies appear in the same order as they do in Table \ref{properties}.  The position angle of the slice is measured eastward of north and included in the top left of each panel. The contours are at -3, -2, 2, 3, 4.5, 6 times the rms in the position-velocity map.  The dashed vertical line corresponds to the optical center of the galaxy.  The dashed horizontal line corresponds to the systemic cz recession velocity.  The dotted horizontal lines delimit the tabulated width of the global HI profile. See the text for more details.}
\label{xvmaps}
\end{figure*}

In addition to the total intensity maps, significance maps, and integrated \hi\ line profiles, in this Appendix we provide images of position-velocity slices along the optical major axis of the galaxy (Figure \ref{xvmaps}).  The images are unique to the moment maps and significance maps because they show the raw data, rather than derived data products.  The galaxies appear in the same order as they are in Table \ref{properties}.  In the top left hand corner of the panels we include the position angle of the optical major axis, which we estimate by-hand measured from north, eastward.  For irregular sources we laid the slice along the largest extent of the \HI\ emission in the column density map.  Contours are drawn in terms of the average noise in the two dimensional maps.

\bsp    
\label{lastpage}
\end{document}